\documentclass[journal]{IEEEtran}

\usepackage{amssymb}
\usepackage{graphicx}
\usepackage{amsmath}
\usepackage{cite}
\usepackage{array, booktabs}
\usepackage{algorithm,algorithmic}
\usepackage{caption}
\newcommand{\br}{{\bf r}}

\newcommand{\bx}{{\bf x}}
\newcommand{\by}{{\bf y}}

\newcommand{\bbb}{{\bf B}}
\newcommand{\bbc}{{\bf C}}
\newcommand{\bbd}{{\bf D}}
\newcommand{\bbe}{{\bf E}}
\newcommand{\bbf}{{\bf F}}
\newcommand{\bbg}{{\bf G}}

\newcommand{\bbi}{{\bf I}}
\newcommand{\bbm}{{\bf M}}
\newcommand{\bbq}{{\bf Q}}

\newcommand{\bbu}{{\bf U}}
\newcommand{\bbx}{{\bf X}}

\newcommand{\bbz}{{\bf Z}}

\newcommand{\bbpsi}{{\boldsymbol\Psi}}
\newcommand{\xm}{{\bf x}_m}

\newcommand{\diag}{\mathrm{diag}}
\newcommand{\os}{\alpha_{\textrm{os}}}
\newcommand{\sinc}{\mathrm{sinc}}

\newcommand{\qedsymbol}{$\blacksquare$}

\def\expl{{e^{-i \frac{2\pi}{T}lt}}}

\begin{document}

\title{Sub-Nyquist SAR via Fourier Domain\\ Range Doppler Processing}
\author{Kfir~Aberman
        and~Yonina~C. Eldar,~\IEEEmembership{Fellow,~IEEE}
\thanks{This project has received funding from the European Union's Horizon 2020 research and innovation program under grant agreement No. 646804-ERC-COG-BNYQ, and from the Israel Science Foundation under Grant no. 335/14.}}
\maketitle

\begin{abstract}
Conventional Synthetic Aperture Radar (SAR) systems are limited in their ability to satisfy the increasing requirement for improved spatial resolution and wider coverage. The demand for high resolution requires high sampling rates, while coverage is limited by the pulse repetition frequency. Consequently, sampling rate reduction is of high practical value in SAR imaging. In this paper, we introduce a new algorithm, equivalent to the well-known Range-Doppler method, to process SAR data using the Fourier series coefficients of the raw signals. We then demonstrate how to exploit the algorithm features to reduce sampling rate in both range and azimuth axes and process the signals at sub-Nyquist rates, by using compressed sensing (CS) tools. In particular, we demonstrate recovery of an image using only a portion of the received signal's bandwidth and also while dropping a large percentage of the transmitted pulses. The complementary pulses may be used to capture other scenes within the same coherent processing interval. In addition, we propose exploiting the ability to reconstruct the image from narrow bands in order to dynamically adapt the transmitted waveform energy to vacant spectral bands, paving the way to cognitive SAR. The proposed recovery algorithms form a new CS-SAR imaging method that can be applied to high-resolution SAR data acquired at sub-Nyquist rates in range and azimuth. The performance of our method is assessed using simulated and real data sets. Finally, our approach is implemented in hardware using a previously suggested Xampling radar prototype.
\end{abstract}


\section{Introduction}
\IEEEPARstart{S}{ynthetic} aperture radar (SAR) is a well proven radar imaging technology that enables the production of high-resolution images of targets and terrain. SAR can be operated at night and in adverse weather conditions, overcoming limitations of optical and infrared systems. The basic idea of SAR is that a single monostatic radar transmits pulses at microwave frequencies at a uniform pulse repetition interval (PRI) as it moves along a path. The echoes coming from ground scatterers are then collected and processed in order to generate a focused image. The coherent information recorded at the different positions is used to synthesize a long antenna in order to improve resolution.

Processing of SAR data requires two-dimensional space-variant correlation of the raw data with the point scatter response of the SAR data acquisition system \cite{SAR1999Soumekh}. A full two-dimensional time domain correlation can handle the space-variance, but is computationally inefficient. In order to accelerate computation time, various algorithms have been developed that impose different approximations on the correlation kernel \cite{bamler1992comparison,raney1994precision}. The Range-Doppler Algorithm (RDA) is the most widely used approach for high resolution processing of SAR data. It is conceptually the simplest, can accommodate range varying parameters and is independent of the transmitted pulse structure. An important part of RDA is the Range Cell Migration Correction (RCMC) operation, which is aimed at decoupling the dependency between the two dimensions of the system, range and azimuth, which are also known as fast-time and slow-time, respectively. This step requires fine delay resolution in the Range-Doppler domain, which is typically obtained by digital interpolation \cite{SARdspCumming}. Interpolation allows to reduce the sampling rate at the cost of additional digital computations which effectively increase the rate in the digital domain. In practice, oversampling is often employed to eliminate artifacts caused by digital implementation of standard RDA processing.

\begin{table*}
\centering
\caption{List of Notation.}
\def\arraystretch{1.1}
\begin{tabular}{c | l }
\hline
$h(t), \;H[l]$ & Transmitted signal and its Fourier coefficients \\
$B_h$ & SAR transmitted signal's bandwidth \\
$f_s$ & Receiver sampling rate\\
$f_c$ & Carrier frequency\\
$T$ & Pulse repetition interval (PRI) \\
$N = \lfloor Tf_s\rfloor$ & Number of samples dictated by receiver sampling rate \\
$M$ & Number of transmitted pulses\\
$\os$ & Time oversampling factor \\
$\sigma(\br)$ & Scene's reflectivity map \\
$\vec{v}$ & SAR platform's speed \\
$c$ & Speed of light \\
$\Theta_a$ & Antenna's angular aperture\\
$d_m(t),\;d[n,m],\; D_m[l]$ & Returned signal from the $m$th pulse: continuous, sampled and the Fourier coefficients \\
$\tilde{d}[n,m],\; \tilde{D}_m[l]$ &  Raw data after range compression and its Fourier coefficients \\
$S_k(t),\:S[n,k],\; S_k[l]$ & Data after azimuth DFT:  continuous, sampled and the Fourier coefficients \\
$C_k(t),\;C[n,k],\; C_k[l]$ & Data after RCMC: continuous, sampled and the Fourier coefficients \\
$Y[n,k]$ & Data after azimuth compression\\
$q_{k,l}(t)$ & Weight function for Fourier coefficients relationship in RCMC  \\
$Q_{k,l}[n]$ & Fourier coefficients of the weight function $q_{k,l}(t)$ \\
$\nu(k,l)$ & Subset of coefficients used for approximation of $C_k[l]$ \\
$\beta_m$ & The set of Fourier coefficients of $d_m(t)$ that correspond to its bandwidth\\
$B$ & Cardinality of  $\beta_m$ \\
$\beta_k$ & The set of Fourier coefficients of $C_k(t)$ that correspond to its bandwidth\\
\hline
\end{tabular}
\label{tab:symbolList}
\end{table*}

According to the Shannon-Nyquist theorem, the minimal sampling rate at the SAR receiver should be at least twice the bandwidth of the detected signal in order to avoid aliasing \cite{eldar2015sampling}. In addition, the need to avoid azimuth ambiguities in the resulting image is translated into a minimal pulse repetition frequency (PRF) requirement. The PRF has to be greater than the Doppler bandwidth of the received signals which is dictated by several system parameters, i.e, platform velocity, carrier frequency and the real antenna aperture. This, in fact, limits the maximal swath of the system \cite{curlander1991synthetic}. Consequently, this two-dimensional dense sampling results in large data rates, requiring large on board memory which may be restricted by downlink throughput requirements, especially for orbital missions.

The emerging theory of compressive sensing (CS) states that a signal which is sparse in some basis, can be reconstructed from highly incomplete samples or measurements \cite{eldar2012compressed,cande2008introduction}. Since a SAR image is a map of a spatial distribution of the reflectivity function of stationary targets and terrain, many SAR images are sparse or compressible under an appropriate basis such as wavelet, curvelet or total variation \cite{samadi2011sparse}. In this paper we show that CS can be applied on both dimensions of SAR. Rate reduction in range is realized by low rate analog-to-digital conversion (ADC) at the receiver and azimuth subsampling is expressed by the transmission of a smaller number of pulses during a coherent processing interval (CPI).

\subsection{Related Work}
CS theory has shown promising results in the field of sub-Nyquist sampling in radar applications. The use of Fourier series coefficients in pulse-Doppler radar enables practical sub-Nyquist sampling when the illuminated scene consists of moving targets that correspond to a sparse range-Doppler map \cite{bar2014sub,baransky2014sub,eldar2015clutter}. CS has also been explored in a wide range of radar imaging applications \cite{potter2010sparsity}. In \cite{alonso2010novel}, the authors applied CS on SAR images by separating the processing into two decoupled one-dimensional operations. They showed that CS theory can then be applied in order to reduce the rate in azimuth. However, since RCMC is ignored, this method does not consider system setups with range varying parameters, hence, the quality of some images might be degraded.

The authors in \cite{fang2014fast} and \cite{dong2014novel} used CS in order to reduce the rate in both dimensions. In \cite{fang2014fast} RDA and CS were combined in order to exploit RDA benefits, however, only linear interpolation was considered. To achieve accurate results, the data is normally oversampled and the kernel of the interpolator may span many samples which comes at the expense of efficiency and computational load. In \cite{dong2014novel}, due to its simplicity, the authors suggest a compressive sensing algorithm based on the chirp scaling algorithm (CSA). This processing technique does not require interpolation \cite{raney1994precision}. Unlike RDA, this method is based on the assumption that the transmitted signal has a chirp form and is known to be less robust to noise. Both methods apply random sampling in time without proposing a practical sampling mechanism which enables the extraction of the low-rate samples directly from the analog signals.

Following subsampling, most of the existing CS imaging schemes stack the entire two-dimensional reflectivity map into a vector in order to apply CS recovery methods. For real SAR images, this vectorization operation results in large memory requirements and long reconstruction times. Alternatively, the authors in \cite{yang2013segmented} suggested to split the image into segments and use several computing units to process the data in parallel and solve the vectorized CS problem. This approach achieves better runtime, but does not utilize the two-dimensional structure of the SAR sampling problem.

\subsection{Contributions}
Our contribution is divided into three parts. First, we present a new algorithm, equivalent to RDA, which handles the burden of time interpolation via Fourier series coefficients. Our approach is based on a technique recently developed for ultrasound imaging, called beamforming in frequency \cite{wagner2012compressed,chernyakova2014fourier}. This method shows that conventional beamforming in time which is used to process ultrasound signals can be equivalently performed in the Fourier domain. Adapting this concept to SAR, the required non-integer non-constant shifts in the RCMC stage are performed in frequency using similar techniques. This leads to a new approach of Fourier domain RDA which is completely equivalent to conventional RDA processing and preserves image integrity. An advantage of this method is that it allows to bypass oversampling which is dictated by digital implementation of conventional RDA.

The second contribution is a two-dimensional sub-Nyquist SAR system. Relying on Fourier domain RDA and CS, our system enables sampling both range and azimuth axes below the Nyquist rate. In the range direction, using the Xampling approach \cite{mishali2011xampling,michaeli2012xampling}, we develop a SAR system that samples with practical low rate ADCs. The Xampling (``compressed sampling") methodology, uses an architecture that includes an ADC which performs analog prefiltering of the signal before taking point-wise low-rate samples in order to generate sub-Nyquist Fourier coefficients within certain bands instead of the entire wideband \cite{SubNyquistMagazine}. In the azimuth direction, the reduction allows to process the data and reconstruct the image when the number of processed pulses during a CPI is lower than that required by Nyquist. When one is interested only in range subsampling, we offer a simplified system with better run time \cite{aberman2016range}.

Our sparse recovery algorithm is performed without the use of vectorization, by exploiting the natural two-dimensional structure of SAR data. The core of the method is based on the fast iterative shrinkage thresholding algorithm (FISTA) \cite{beck2009fast,palomar2010convex}, which allows to handle practical limitations of real SAR data. Using various sparsifying transforms, simulations provided in Section~\ref{sec:SimulationResults} show that following reduction of $24\%$ of the Nyquist samples in range, our sub-Nyquist sampling and recovery methods preserve classic RDA processing quality. Moreover, a reduction of more than $50\%$ of the transmitted pulses is presented via simulation for the azimuth axis sub-Nyquist sampling. Simultaneous two-dimensional sub-Nyquist sampling is applied on real SAR data of RADARSAT-1 satellite, leading to a total reduction of about $50\%$ of the original samples processed by conventional systems. Along with software simulations, our hardware prototype demonstrates that our technique can cope with practical limitations and fits real radar imaging systems.

Finally, we show how the sub-Nyquist property of our system can be exploited for cognitive SAR and reduced time-on-scene. Specifically, we rely on the basic idea that if we are able to reconstruct the image while sampling only part of the data, then only this part should be transmitted. Thus, we do not have to transmit the whole signal's bandwidth nor the number of pulses required by Nyquist. Consequently, time gaps (during CPI) and frequency holes (within the signal's energy) exist in our system. For azimuth subsampling, analogously to the reduced time-on-target concept applied to radar signals in \cite{cohen2016RToT}, we propose exploiting these time gaps to transmit pulses to another zone, using electronic beam steering. This enables capturing several scenes during the same CPI.
For range subsampling, we focus on adaptive transmission and reception by modifying the emitted signal to transmit only over a small number of narrow frequency bands and use our sparse recovery method. Complying with the concept of cognitive radar (CR) \cite{haykin2005cognitive}, which is defined as a radar system in which both the transmitter and receiver are able to dynamically adjust to the environment conditions, the bands support may vary with time to allow for dynamic and flexible adaptation to the existing spectrum. Such a system allows to cope with overloaded spectrum by using a smaller portion of it. In addition, by concentrating all the available power in the transmitted narrow bands rather than over a wide spectral band, we increase the signal to noise ratio (SNR)  \cite{cohen2016RToT}. The fact that we earn higher coverage and better SNR by exploiting the missing data, leads to a sub-Nyquist SAR system which outperforms conventional systems.

The remainder of this paper is organized as follows: In Section~\ref{sec:SARmodel} we describe the SAR model, the assumptions we use for its simplification and review the classic Range-Doppler algorithm. In Section~\ref{sec:FourierRCMC} we introduce the Fourier domain Range-Doppler method. Our two-dimensional sub-Nyquist system using Fourier domain RDA is described in Section~\ref{sec:2dsubNyquist}, along with an analysis of noiseless recovery. In Section~\ref{sec:gaps} we introduce our cognitive and reduced time-on-scene SAR systems. Simulation results on simulated and real data are presented in Section~\ref{sec:SimulationResults}. Finally, we show how our approach is integrated into a stand-alone system, using National Instrument (NI) hardware.
Table~\ref{tab:symbolList} summarizes the important notation used throughout the paper.

\section{SAR Model and the Range-Doppler Algorithm}
\label{sec:SARmodel}
SAR spaceborne and airborne systems are based on a radar which travels along a well defined path with velocity $\vec{v}$ and transmits every PRI, $T$, a time-limited pulse $h(t)$ with negligible energy at frequencies beyond $B_h/2$. The transmitted pulses are sent from $M$ different locations, $\left\{\xm\right\}_{m=0}^{M-1}$, where ${\bf x}_{0}$ is the origin and $\left\|\xm - {\bf x}_{0}\right\| = m\left|\vec{v}\right| T$ is the platform displacement at the $m$th location. The pulses are transmitted into a scene with a stationary terrain reflectivity, $\sigma(\br)$, where $\br=(x,r)$ is the scene spatial vector consisting of azimuth and range axes, respectively.

\begin{figure}
\centering
 \includegraphics[width=\linewidth]{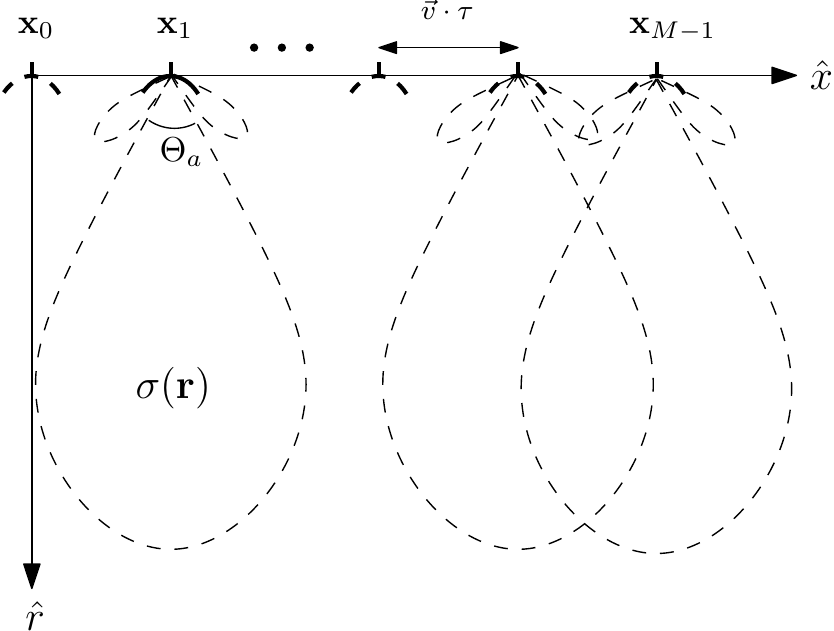}
\caption{SAR system model for the stripmap mode. The coherent information is recorded at different positions, spaced by a displacement of  $\vec{v}\tau$. At every position the radar, which has an angular aperture of $\Theta_a$,  captures a part of the scene reflectivity, $\sigma(\br)$. }
\label{fig:SARmodel}
\end{figure}

The pulse $h(t)$ is modulated by a pure tone with carrier frequency $f_c$, so that the transmitted signal is $h(t)e^{j2\pi f_ct}$.
The received signal from the $m$th transmitted pulse, after coherent demodulation, is given by
\begin{eqnarray}
\label{eq:received}
	d_m(t) & = & \int\sigma(\br)h(t-2\left\|\br-\xm\right\|/c)w_a(\xm,\br)\times\\\nonumber
	& & e^{-j4\pi f_c \left\|\br-\xm\right\|/c}d\br,
\end{eqnarray}
where $\left\|\br-\xm\right\|$ is the distance from the radar at position $\xm$ to a scatter point at position $\br$ (no sensor movement is assumed between transmission and reception of a pulse -- the ``stop-and-hop'' assumption) and $w_a(\xm,\br)$ is the antenna beam pattern. The beam generally forms a spatial squared sinc function with an angular aperture (main lobe) of $\Theta_a$ that is inversely proportional to the antenna length. Its steering direction varies depending on the SAR operation mode (stripmap, spotlight, scan SAR, etc.), which are mainly distinguished by resolution and coverage capabilities \cite{SARdspCumming}. For the stripmap mode, the beam pattern is
\begin{equation}
	w_a(\xm,\br) = \sinc^2\left(\frac{\left|x - x_m\right|}{r}{\cot\frac{\Theta_a}{2}}\right),
\label{eq:beampattern}
\end{equation}
where $x_m$ denotes the azimuth coordinate, $\hat{x}$, of $\bx_m$.
A SAR system model, for the stripmap mode, is depicted in Fig.~\ref{fig:SARmodel}.
In practice, $d_m(t)$ will be contaminated by additive white Gaussian noise.

The goal of SAR imaging is to reconstruct the complex scene reflectivity, $\sigma(\br)$, from the raw data returns in \eqref{eq:received}. In order to perform processing, the analog signals are first sampled. According to the Nyquist theorem, $d_m(t)$ should be sampled at least at $B_h$, creating $d[n,m]=d_m(nT_s)$, with $0\leq n < N =\left\lfloor Tf_s\right\rfloor$, where $f_s=1/T_s$ is the sampling rate at the receiver. In addition, the need to avoid azimuth ambiguities in the resulting radar image leads to the requirement of dense spatial sampling of the entire scene. This dense sampling results in a minimum PRF requirement, which spatially samples the scene every PRI. In the stripmap mode an approximation of the beam pattern in \eqref{eq:beampattern} to a time window, leads to a Doppler bandwidth of $2v/l_a$, where $l_a$ is the actual antenna length \cite{SARdspCumming}.

After sampling the data is processed. Since the SAR acquisition system is not space-invariant, various algorithms have been developed in order to approximate the reflectivity $\bbi\approx\sigma(\br)$ and to accelerate processing time \cite{raney1994precision}. RDA is the most common approach and has one of the best accuracy/generality/efficiency tradeoffs among existing algorithms \cite{bamler1992comparison}. There are three main steps in implementing RDA:
\begin{enumerate}
	\item Range compression.
	\item Range cell migration correction.
	\item Azimuth compression.
\end{enumerate}

In order to simplify the mathematical expressions we use the ``low squint angle'' assumption in our derivation, namely, the angle between the normal of the antenna's plane and the direction of transmission is assumed to be small. This means that secondary range compression (SRC) is not used in the processing flow. For high-squint cases, we can incorporate SRC as another linear operator and modify the azimuth matched filter accordingly, in order to enhance the focusing ability \cite{SARdspCumming}.

The range compression stage uses the pulse compression property which states that $h(t)\ast h^{\ast}(-t)=\delta(t)$, where $\delta(t)$ is a narrow pulse with width $1/B_h$. The raw data $d[n,m]$ is therefore compressed in the range direction to
\begin{equation}
	\tilde{d}[n,m] = d[n,m]\ast h^{\ast}[-n].
\label{eq:proc}
\end{equation}
Next, the raw data is transformed to the range-Doppler domain via the discrete Fourier transform (DFT) along the azimuth axis:
\begin{equation}
	S[n,k] = \text{DFT}_m\left\{\tilde{d}[n,m]\right\}=\sum_{m=0}^{M-1}\tilde{d}[n,m]e^{-j2\pi km/M},
\label{eq:procRD}
\end{equation}
followed by RCMC. The purpose of RCMC is to compensate for the effect of range cell migration which were migrated from their origin due to the varied satellite-scatterer distance and to correct the hyperbolic behavior of the target trajectories. The RCMC operator can be written as
\begin{equation}
	C[n,k] = S\left[n + n\cdot ak^2 ,k\right].
\label{eq:RCMC}
\end{equation}
For every Doppler frequency $k$, the range axis is scaled by $1 + ak^2$. In stripmap mode we have, for example, $a = \frac{\lambda^2}{8{\left|\vec{v}\right|}^2T^2M^2}$.
As can be seen in \eqref{eq:RCMC}, this range-variant shift requires values which fall outside the discrete grid.
There are two ways to implement RCMC: In the first option, RCMC is performed by range interpolation in the Range-Doppler domain. However, this interpolation is time-consuming and computationally demanding. The second approach involves the assumption that the range cell migration is range invariant, at least over a finite range block. In this case, RCMC is implemented using an DFT, linear phase multiply, and inverse DFT (IDFT) per block. However, this implementation has the disadvantage that blocks have to overlap in range, and the efficiency gain may not be worth the added complexity.

\begin{figure*}
\centering
\begin{tabular}{ccccc}
\hspace{-10mm} \includegraphics[width=0.19\linewidth]{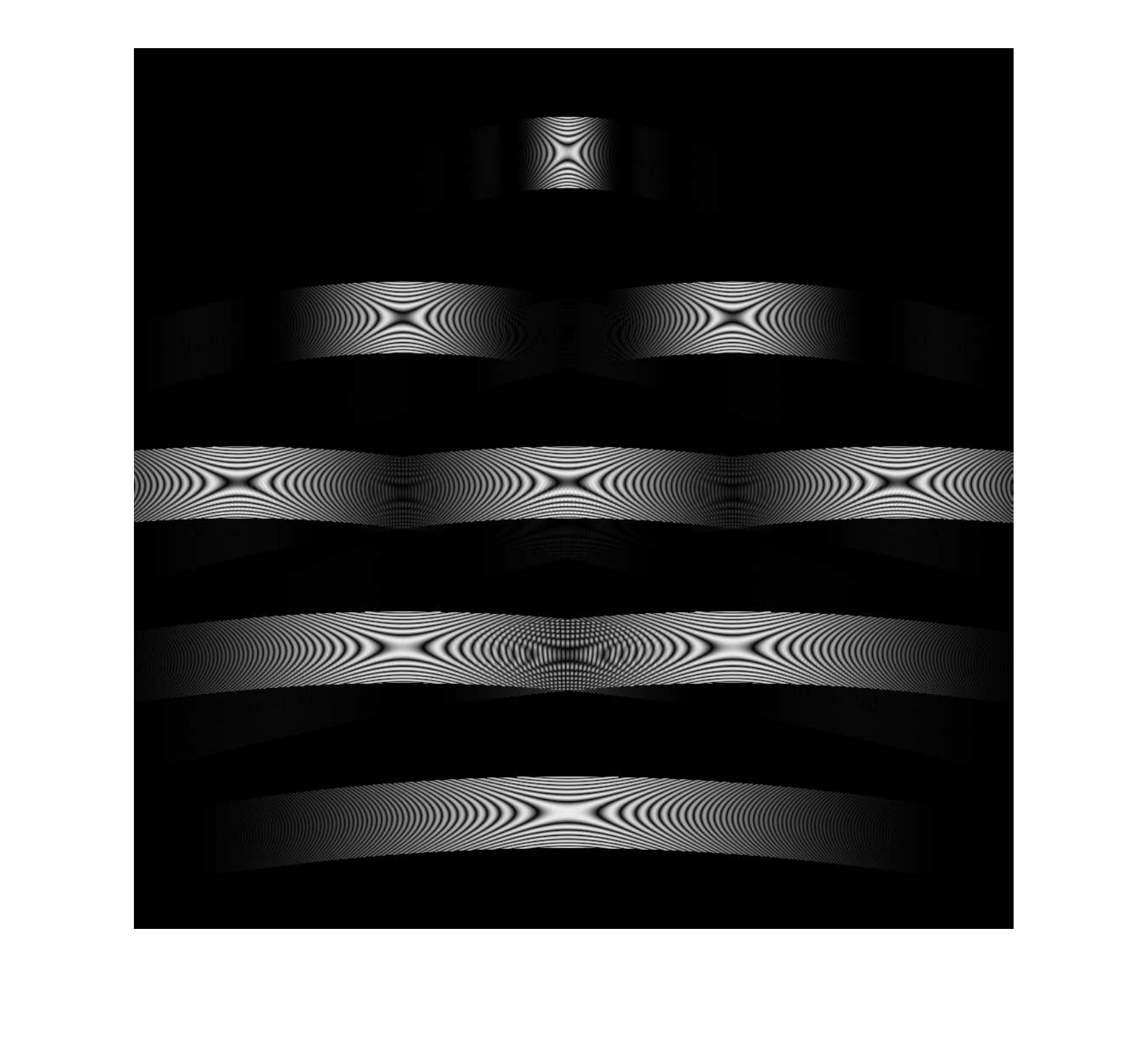} & \includegraphics[width=0.19\linewidth]{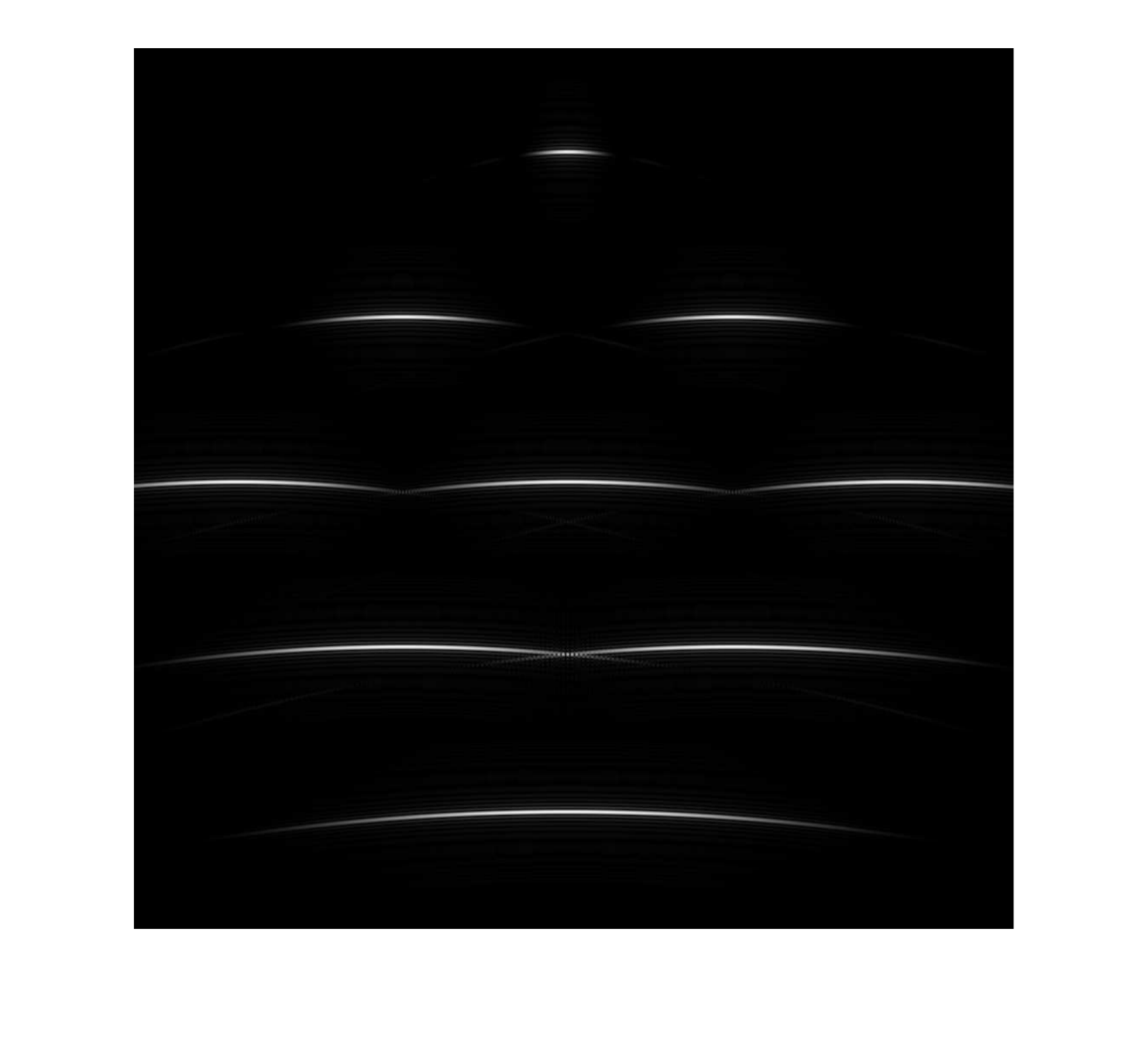} & \includegraphics[width=0.19\linewidth]{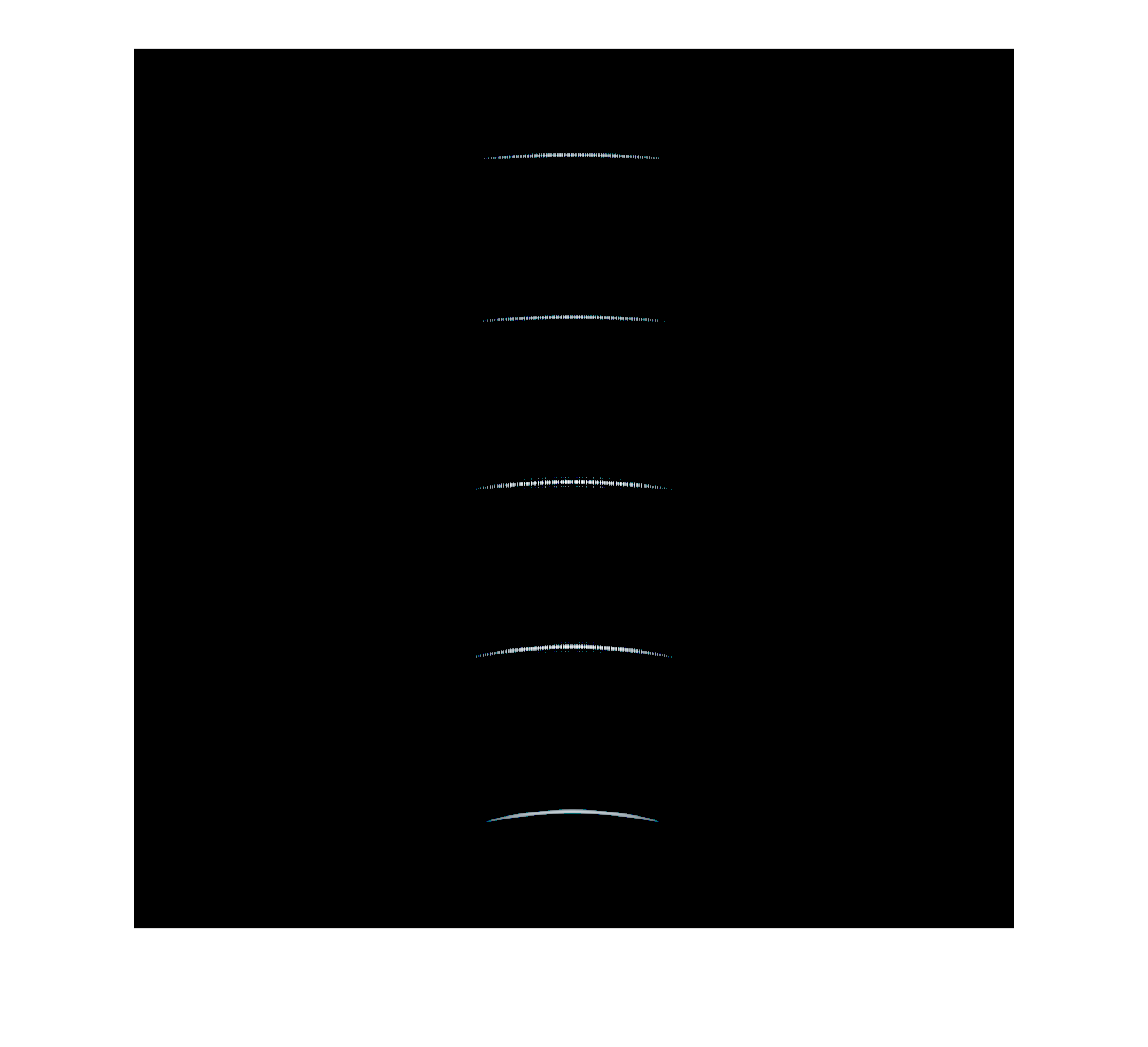} & \includegraphics[width=0.19\linewidth]{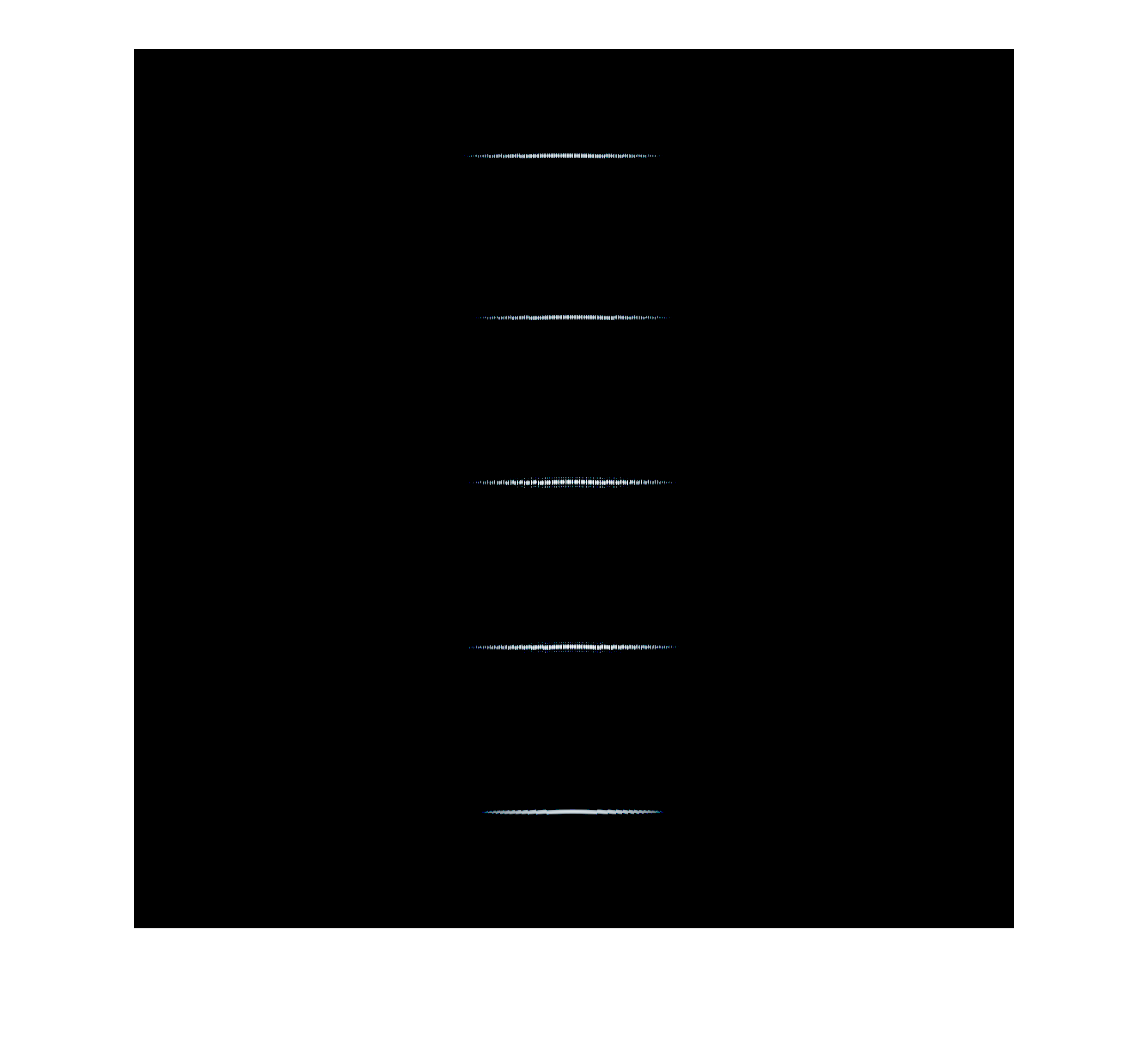} & \includegraphics[width=0.19\linewidth]{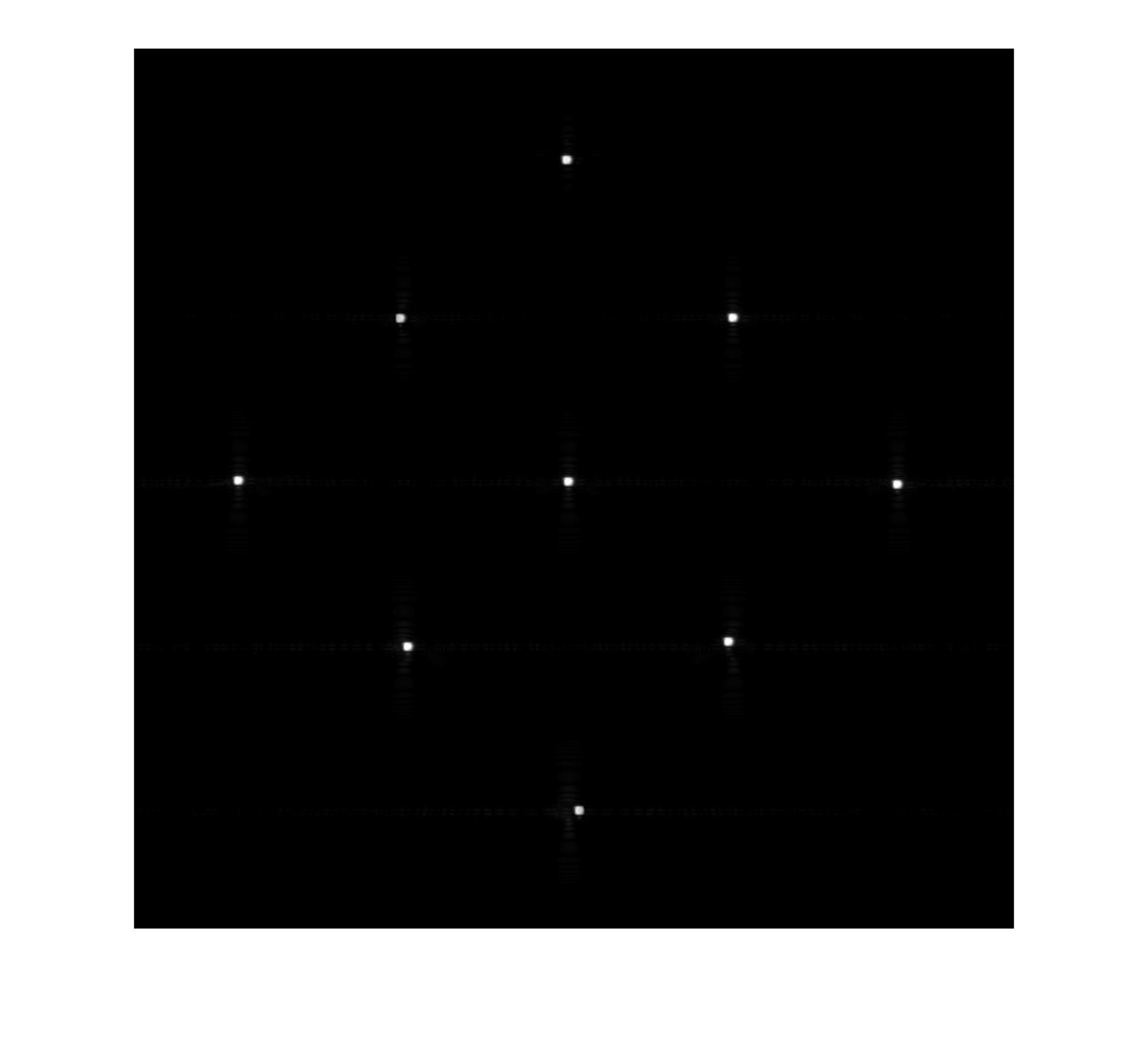} \\
\hspace{-10mm} (a) & (b) & (c) & (d) & (e)\\
\end{tabular}
\caption{The Range-Doppler Algorithm (RDA) stages for equally spaced single-point reflectors. (a) Raw data (real component). (b) Range compression. (c) Azimuth DFT. (d) Range cell migration correction (RCMC). (e) Compressed image. }
\label{fig:RDA}
\end{figure*}

Following RCMC, the signal is compressed in the azimuth direction. The low squint angle assumption enables the compression by using a matched filter of a linear chirp \cite{SARdspCumming}:
\begin{equation}
	Y[n,k] = C[n,k]e^{-j\pi\frac{k^2}{K_a[n]}},
\label{eq:azimuthCompression}
\end{equation}
where $K_a[n]$ is the range dependent azimuth chirp rate
\begin{equation}
	K_a[n] = \frac{4M^2T^2{\left|\vec{v}\right|}^2}{\lambda cnT_S}.
\label{eq:azimuthChirpRate}
\end{equation}
An IDFT in the azimuth direction results in the focused data:
\begin{equation}
I[n,m] = \text{IDFT}_k\left\{Y[n,k]\right\}=\frac{1}{M}\sum_{k=0}^{M-1}Y[n,k]e^{j2\pi mk/M}.
\label{eq:IFFTazimuth}
\end{equation}
Figure~\ref{fig:RDA} depicts the RDA stages for equally spaced single-point reflectors.

RDA is the preferred algorithm in most SAR operations thanks to its high precision and generality. However, its main disadvantage is the increase in processing due to the extra interpolation. Thus, processing in the time domain imposes a high sampling rate and considerable burden on the RCMC block. We next show that the number of
samples can be reduced significantly by exploiting ideas of processing in the Fourier domain, sub-Nyquist sampling and CS-based signal reconstruction.

\section{Range Doppler Algorithm via Fourier Coefficients}
\label{sec:FourierRCMC}
In this section we show that RDA can be performed in frequency, using the Fourier series coefficients of the raw data, paving the way to substantial reduction in the number of samples needed to obtain the same image quality. In particular, we adapt the idea of compressed beamforming in ultrasound imaging \cite{chernyakova2014fourier,wagner2012compressed}, to perform RCMC using Fourier series coefficients instead of the expensive time-domain interpolation without any assumptions on the signal structure or the invariance of range blocks. This allows to transfer the process of RDA to the frequency domain, and eliminate the need for oversampling.

\subsection{Fourier Domain RCMC}
\label{subsec:FourierDomainRCMC}
Similarly to \cite{chernyakova2014fourier} we begin by calculating the Fourier series coefficients of the continuous version of \eqref{eq:RCMC}
\begin{equation}
	C_k(t) = S_k\left(t(1 + ak^2)\right),
\label{eq:RCMCcont}
\end{equation}
where $S_k\left(nT_s\right)=S[n,k]$. Denote the Fourier series coefficients of $C_k(t)$ with respect to the interval $[0,T)$ by
\begin{equation}
	C_k[l]=\frac{1}{T}\int_0^T I_{[0,T_k)}(t)C_k(t)\expl dt,
\label{eq:correctedRangeFourierCoeff}
\end{equation}
where $T_k=T/(1+ak^2)$ and $I_{[a,b)}$ is the indicator function which equals $1$ when $a \leq t < b$ and $0$ otherwise. Substituting \eqref{eq:RCMCcont} into \eqref{eq:correctedRangeFourierCoeff} we get
\begin{equation}
\label{eq:ckl}
	C_k[l]=\frac{1}{T}\int_0^T S_k(t)q_{k,l}(t) dt,
\end{equation}
with
\begin{equation}
\label{eq:qkl}
	q_{k,l}(t) =  I_{[0,T)}(t) \frac{1}{1+ak^2}\exp\left\{{-i\frac{2\pi}{T}lt\left(\frac{1}{1+ak^2}\right)}\right\}.
\end{equation}

We next express $S_k(t)$ in terms of its Fourier series coefficients representation
\begin{equation}
\label{eq:SkFourierRep}
S_k(t)=\sum_{n=-\infty}^{\infty}S_k[n]e^{i\frac{2\pi}{T}nt}.
\end{equation}
Substituting into \eqref{eq:ckl} leads to
\begin{align}
\label{eq:cklsum}
C_k[l] & = \frac{1}{T}\int_0^T\sum_{n=-\infty}^{\infty}S_k[n]e^{i\frac{2\pi}{T}nt}q_{k,l}(t) dt \\\nonumber
			 & =  \sum_{n=-\infty}^{\infty}S_k[n]\frac{1}{T}\int_0^Tq_{k,l}(t)e^{-i\frac{2\pi}{T}(-n)t}dt \\\nonumber
			 & =  \sum_{n=-\infty}^{\infty}S_k[n]Q_{k,l}[-n],
\end{align}
where $Q_{k,l}[n]$ are the Fourier series coefficients of $q_{k,l}(t)$.
Using the relationship between the continuous time Fourier transform (CTFT) $X(\omega)$ and the Fourier series coefficients $C[l]$ of a finitely supported function $x(t)$, $C[l] = \frac{1}{T}X\left(\frac{2\pi}{T}l\right)$, we get
\begin{equation}
Q_{k,l}[n]=\frac{1}{1+ak^2}e^{-j\pi\left(n+\frac{l}{1+ak^2}\right)}\sinc\left(n+\frac{l}{1+ak^2}\right).
\label{eq:Qs}
\end{equation}

\begin{figure}
\centering
\includegraphics[width=\linewidth]{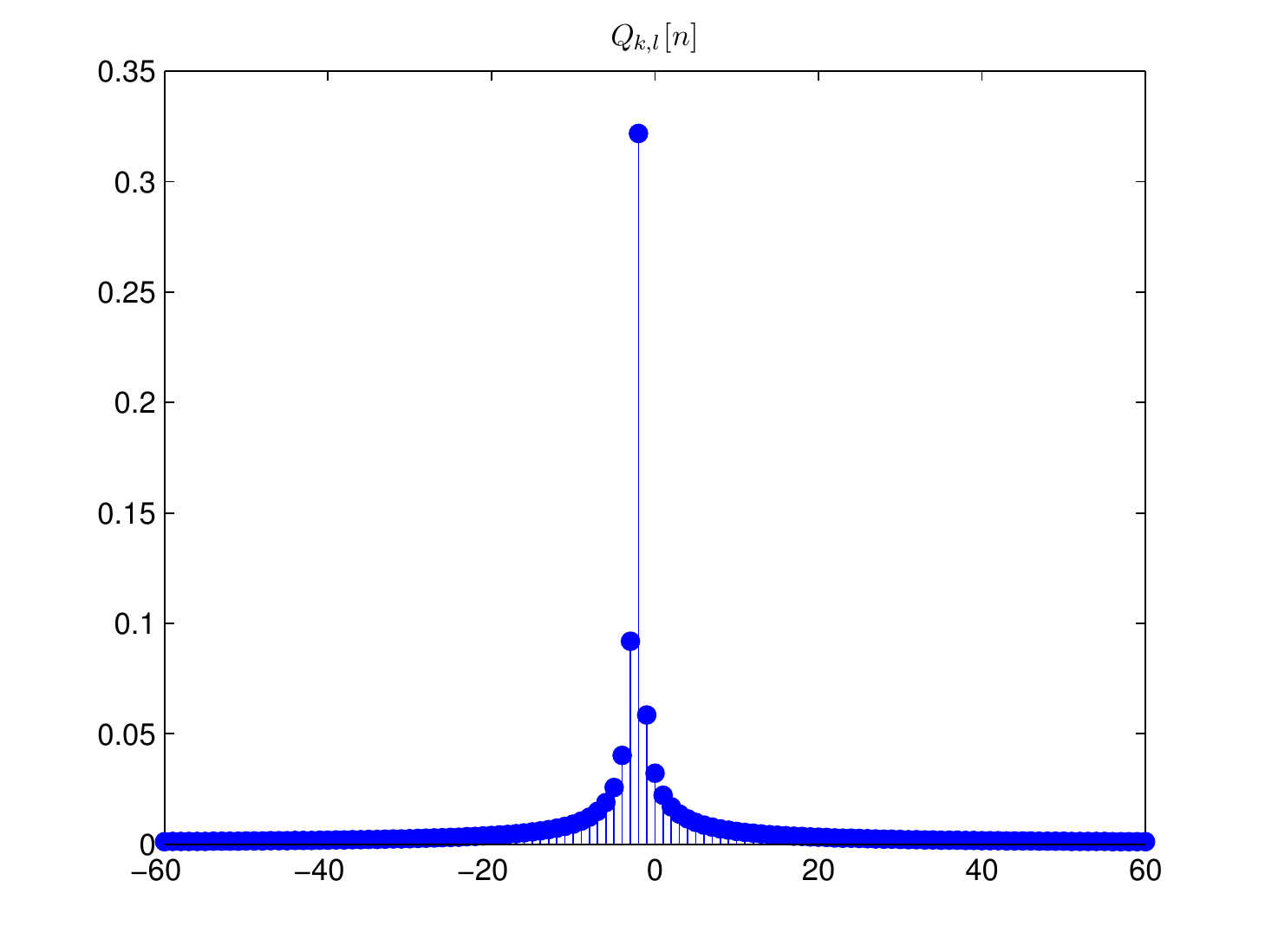}
\caption{The Fourier series coefficients $\left\{Q_{k,l}[n]\right\}$ of $q_{k,l}(t)$ are characterized by a rapid decay, where most of the energy is concentrated around $n_{k,l}$. Here $k=4$, $l=4$ and $a=2$, so that $n_{k,l}=-2$.}
\label{fig:fourierCoeffEnergy}
\end{figure}

It is easy to see that most of the energy of the set $Q_{k,l}[n]$ is concentrated around a specific component, $n_{k,l} = \textrm{round}\left(-\frac{l}{1+ak^2}\right)$, where the $\textrm{round}(\cdot)$ operation rounds the argument to its closest integer. This behavior is typical to any choice of $k$ or $l$. An example with $k=4$, $l=4$ and $a=2$ is shown in Fig. \ref{fig:fourierCoeffEnergy}.
Thus, for every Doppler frequency $k$, the Fourier series coefficients of the scaled signal, $C_k(t)$, can be calculated as a linear combination (weighted sum) of a local choice of Fourier series coefficients of $S_k(t)$,
\begin{equation}
\label{eq:fourierRCMC}
C_k[l]=\sum_{n\in\nu(k,l)}S_k[n]Q_{k,l}[-n],
\end{equation}
where $\nu(k,l)$ is the set of indices which is dictated by the decay property of \eqref{eq:Qs}.

We conclude that given $S_k[l]$, \eqref{eq:fourierRCMC} provides the Fourier series coefficients, $C_k[l]$, of the corrected signal $C_k(t)$ defined in \eqref{eq:RCMCcont}.

\subsection{Acquisition in Fourier Domain}
Assuming that the samples can be extracted directly in the Fourier domain, the samples are defined via
\begin{equation}
	D_m[l]=\frac{1}{T}\int_0^Td_m(t)e^{-i\frac{2\pi}{T}lt} dt.
	\label{eq:cmn}
\end{equation}
We next show how the preliminary stages of RDA can be performed in the Fourier domain as well.

Range compression is simply applied in the Fourier domain by
\begin{equation}
	\tilde{D}_m[l]=T\cdot D_m[l]H^{\ast}[l],
	\label{eq:fourierDomainRangeCompression}
\end{equation}
where $H[l]$ is the $l$th Fourier series coefficient of the transmitted pulse, $h(t)$.
Next, we perform azimuth DFT on the range Fourier samples
\begin{equation}
	S_k[l] = \text{DFT}_m\left\{\tilde{D}_m[l]\right\}=\sum_{m=0}^{M-1}\tilde{D}_m[l]e^{-i2\pi km/M}.
	\label{eq:azimuthDFTonFourier}
\end{equation}
For every Doppler frequency we then use \eqref{eq:fourierRCMC} to apply RCMC and calculate the (range) scaled signal Fourier series coefficients. Applying an inverse Fourier transform on $\left\{C_k[l]\right\}$ reconstructs the corrected sampled signal after RCMC,
\begin{equation}
	C[n,k] = \sum_{l=-\infty}^{\infty}C_k[l]e^{i\frac{2\pi}{T}lnT_s}.
	\label{eq:scaledSigReconstruction}
\end{equation}
We then continue to the original procedure by applying \eqref{eq:azimuthCompression} and \eqref{eq:IFFTazimuth} to complete the processing. A comparison between Fourier domain RDA and conventional RDA is introduced in Table \ref{tab:comprasionRDA}.

\begin{table*}[t]
  \centering
  \bgroup
  \def\arraystretch{2.5}
  \begin{tabular}{|c||c||c|}
	\hline
	\textbf{Algorithm stage} &
  \textbf{Range-Doppler Algorithm} & \textbf{Fourier domain RDA}\\
  \hline\hline
   Range compression & $\tilde{d}[n,m] = d[n,m]\ast h^{\ast}[-n]$ & $\tilde{D}_m[l]=T\cdot D_m[l]H^{\ast}[l]$ \\\hline
    Azimuth DFT & $S[n,k] = \sum\limits_{m=0}^{M-1}\tilde{d}[n,m]e^{-j2\pi km/M}$ & $S_k[l] = \sum\limits_{m=0}^{M-1}\tilde{D}_m[l]e^{-j2\pi km/M}$ \\\hline
     RCMC & $C[n,k] = S\left[n + n\cdot ak^2 ,k\right]$ & $C_k[l]=\sum\limits_{n\in\nu(k,l)}S_k[n]Q_{k,l}[-n]$\\\hline
   Azimuth compression & $Y[n,k] = C[n,k]e^{-j\pi\frac{k^2}{K_a[n]}}$ & $Y[n,k] = \left(\sum\limits_{l\in\beta_k}C_k[l]e^{i\frac{2\pi}{T}lnT_s} \right) e^{-j\pi\frac{k^2}{K_a[n]}}$\\\hline
    Azimuth IDFT & $I[n,m] =\frac{1}{M}\sum\limits_{k=0}^{M-1}Y[n,k]e^{j2\pi mk/M}$ & $I[n,m] =\frac{1}{M}\sum\limits_{k=0}^{M-1}Y[n,k]e^{j2\pi mk/M}$\\\hline
  \end{tabular}
  \egroup
  \vspace{3mm}\caption{Fourier domain RDA compared to conventional RDA}
  \label{tab:comprasionRDA}
\end{table*}

\subsection{Sampling and processing at the Nyquist rate}
In practice, SAR signals are sampled at rates which are higher than the Nyquist rate. Moreover, prior to RCMC a subsequent digital interpolation increases the effective rate of the entire system even more. A typical oversampling factor of 1.5 to 4 times the transmitted signal bandwidth is usually used in order to eliminate artifacts caused by digital implementation. While achieving the same results, we next show how our algorithm may be performed without oversampling.

Denote by $\beta_m$, $|\beta_m| = B$, the set of Fourier series coefficients of the detected signal, $d_m(t)$, that correspond to its bandwidth, namely, the values of $l$ for which $D_m[l]$ is nonzero (or larger than a threshold). The ratio between the cardinality of the set $\beta_m$ and the overall number of samples $N = \left\lfloor Tf_s\right\rfloor$ required by standard RDA is dictated by the oversampling factor.

\begin{figure}
\centering
 \includegraphics[width=\linewidth]{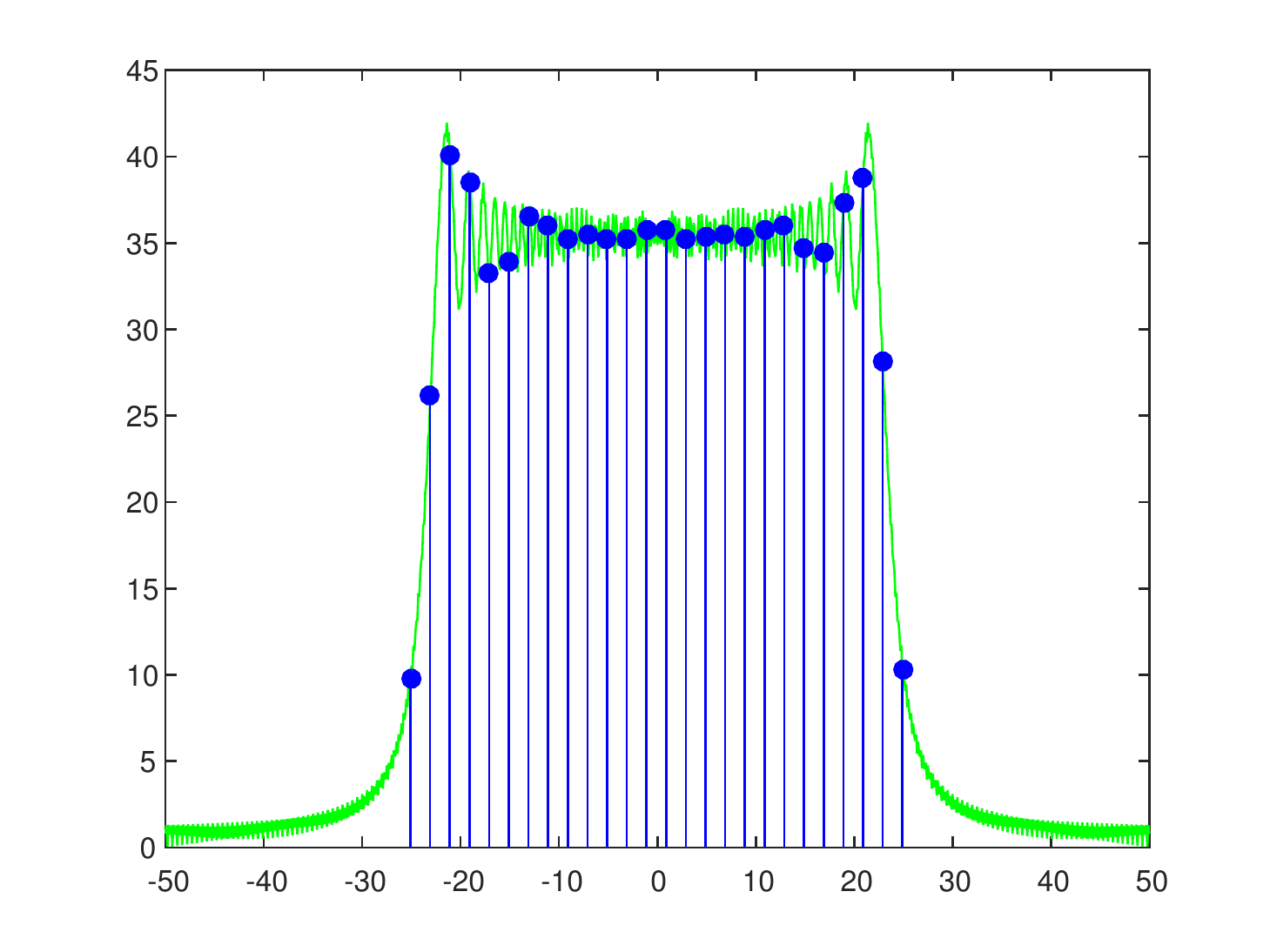}
\caption{Fourier series coefficients are taken within the effective bandwidth of the returned signals. Since the signals are finite in time the coherent information within the discrete frequency samples is sufficient to handle SAR processing. }
\label{fig:effectiveBW}
\end{figure}

The bandwidth of the returned signals in \eqref{eq:received} is equal to the bandwidth of the transmitted signal, $H[l]$. Thus, following range compression in \eqref{eq:fourierDomainRangeCompression}, the bandwidth of the signals remains the same. Moreover, it is easy to see that the azimuth DFT stage in \eqref{eq:azimuthDFTonFourier} preserves the bandwidth of the range compressed signal, and that for every Doppler frequency $k$, the cardinality of the non-zero $\left\{S_k[l]\right\}$ equals to $B$. However, \eqref{eq:fourierRCMC} implies that the bandwidth of the corrected signals following RCMC, $\beta _{k}$, will contain at most $B+\left|\nu(k,l)\right|$ nonzero frequency components. Due to the azimuth DFT operation, to compute the elements in $\beta _{k}$ all we need is the set $\beta_m$ from each one of the detected signals. In a typical imaging setup $B$ is on the order of thousands of coefficients, while $\nu(k,l)$, defined by the decaying properties of $\{Q_{k,l}[n]\}$, is typically no larger than $10$. This implies that $B\gg \left|\nu(k,l)\right|$, so $|\beta_k| = B+\left|\nu(k,l)\right|\approx B$. Hence, the bandwidth of the corrected signals is approximately equal to the bandwidth of the detected signals, which means that sampling and processing can be done at the Nyquist rate and no oversampling is required. In a typical system setup this reduction leads to $B/N = 2/3$ to $1/4$. Figure~\ref{fig:effectiveBW} depicts the Fourier series coefficients which are taken within the effective bandwidth.


\subsection{Simulation and validation}
\label{subsec:SimulationAndValidation}
To demonstrate the equivalence of RDA in time and frequency, we applied both methods on simulated SAR raw data. 

First, we evaluate the required number of $\left|\nu(k,l)\right|$ by measuring the reconstruction quality via the peak sidelobe ratio (PSLR) of the point spread function (PSF) of the system. Assuming that no windowing is applied to the range and azimuth signals, the PSF can be approximately described as a two-dimensional sinc function, and its beam widths in range and azimuth are inversely proportional to the transmitted signal bandwidth and the Doppler bandwidth, respectively. PSLR is defined as the ratio of the peak intensity of the most prominent sidelobe to the peak intensity of the main lobe, i.e., the smaller the PSLR, the better an image quality.

\begin{figure*}
\centering
\begin{tabular}{ccc}
\hspace{-5mm} \includegraphics[width=0.33\linewidth]{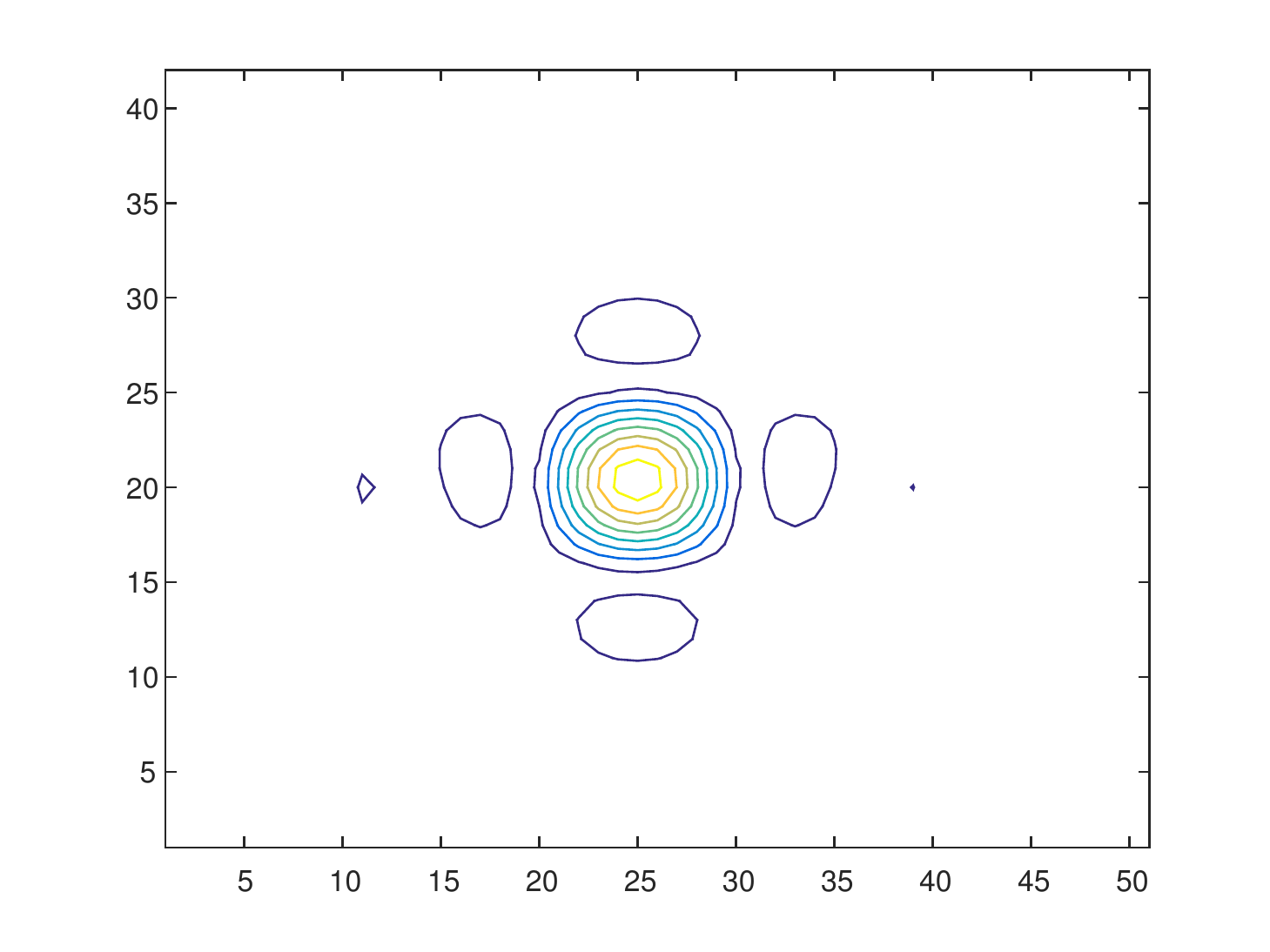} & \includegraphics[width=0.33\linewidth]{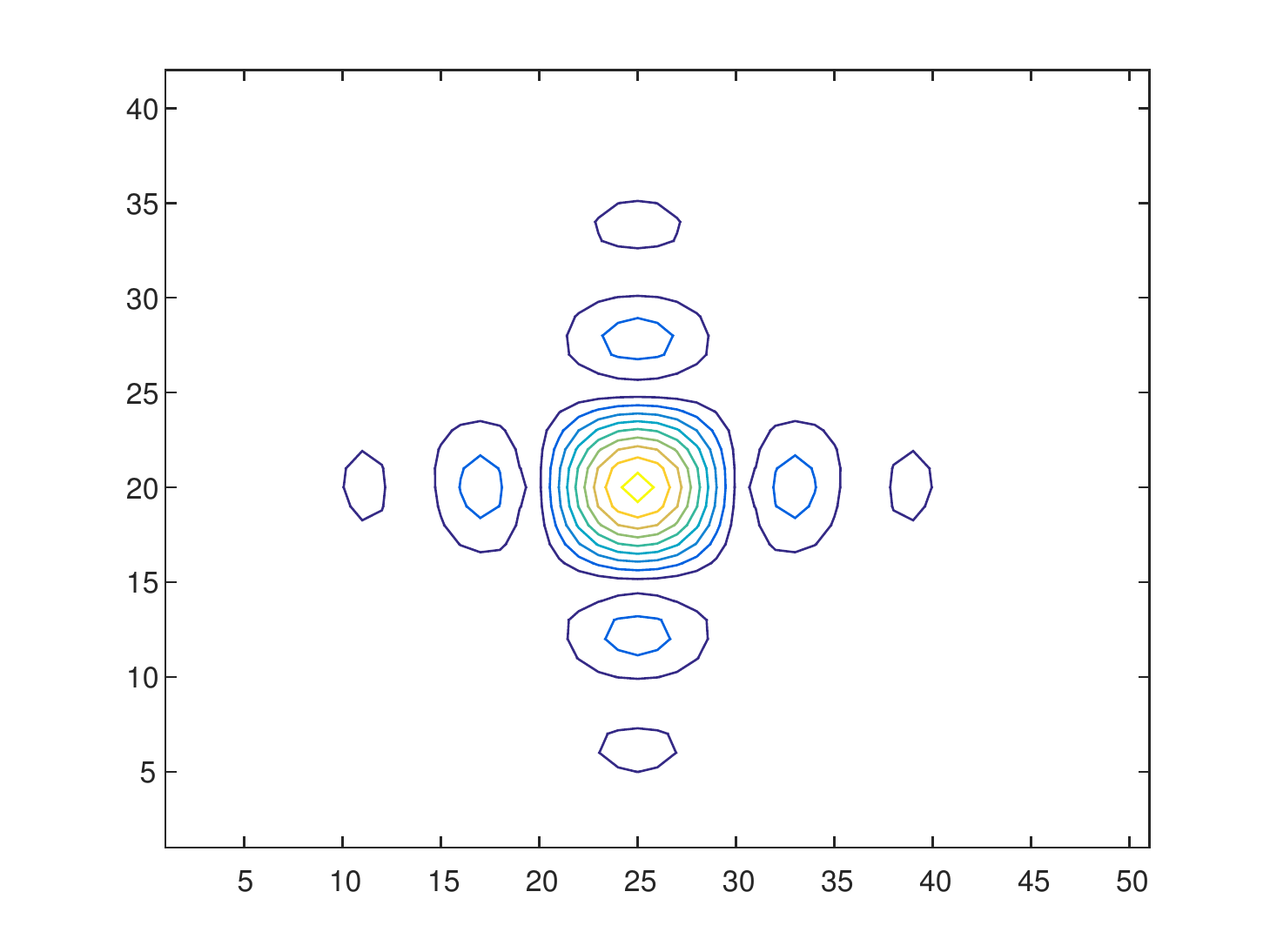} & \includegraphics[width=0.33\linewidth]{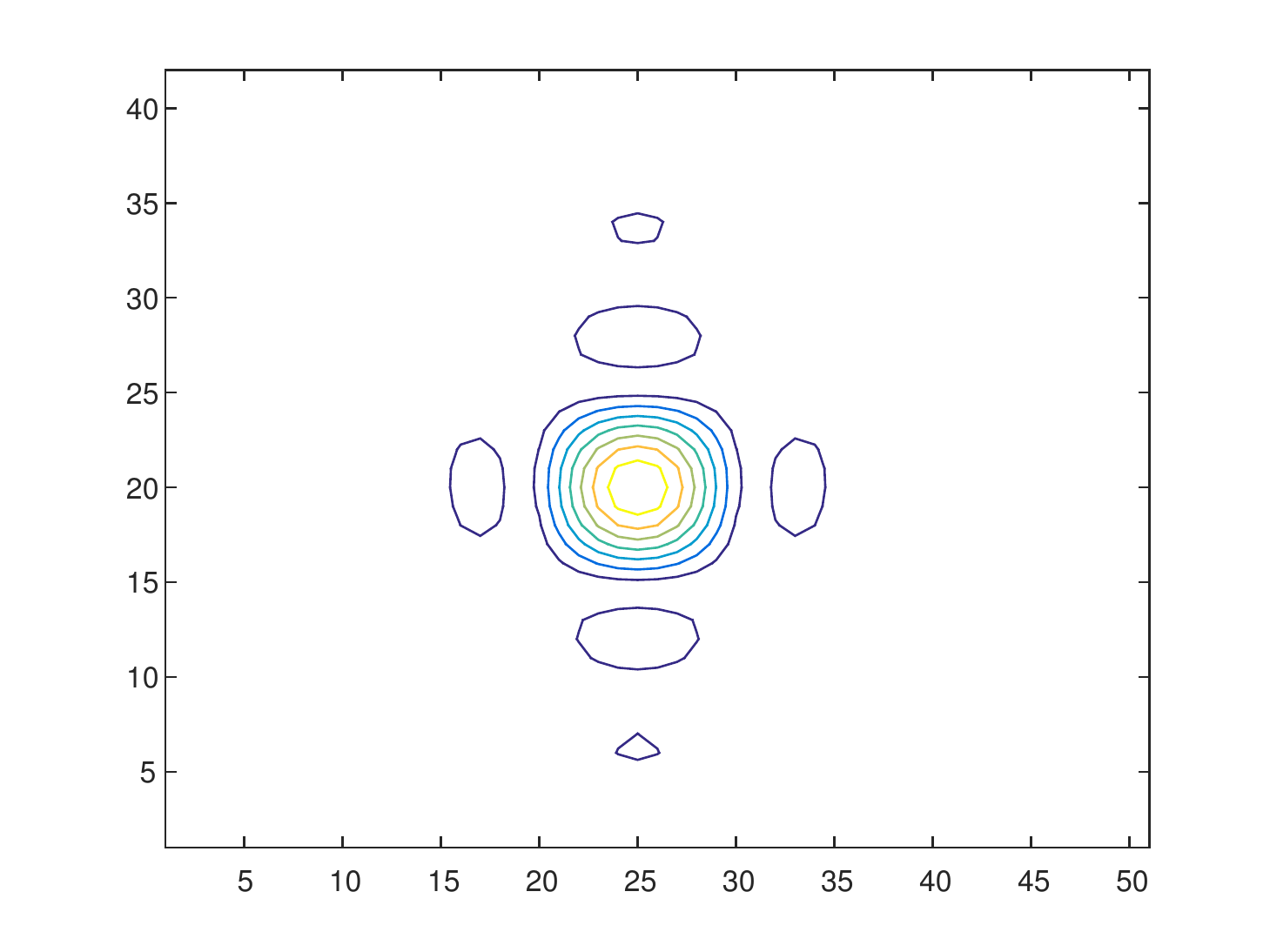} \\
(a) & (b) & (c)\\
\end{tabular}
\caption{A point spread function (PSF) of a SAR system simulated with different RDA techniques (range is the vertical axis and azimuth is the horizontal one). (a) Conventional RDA, PSLR = 13.32 [dB]. (b) Fourier domain RDA, $\left|\nu(k,l)\right| = 3$, PSLR = 11.7598 [dB]. (c) Fourier domain RDA, $\left|\nu(k,l)\right| = 5$, PSLR = 13.29 [dB].}
\label{fig:RCMCtimeVsFourier}
\end{figure*}

The PSF was generated by a single reflector in the scene center as the input of the system, $\sigma(\br) = \delta(\br - \br_c)$. Results are shown in Fig.~\ref{fig:RCMCtimeVsFourier}. It can be seen that $\left|\nu(k,l)\right|=5$ components of each $S_k[n]$ are sufficient to achieve almost the same quality, visually and quantitatively.

\begin{figure}
\centering
\begin{tabular}{cc}
\hspace{-3mm} \includegraphics[width=0.98\columnwidth]{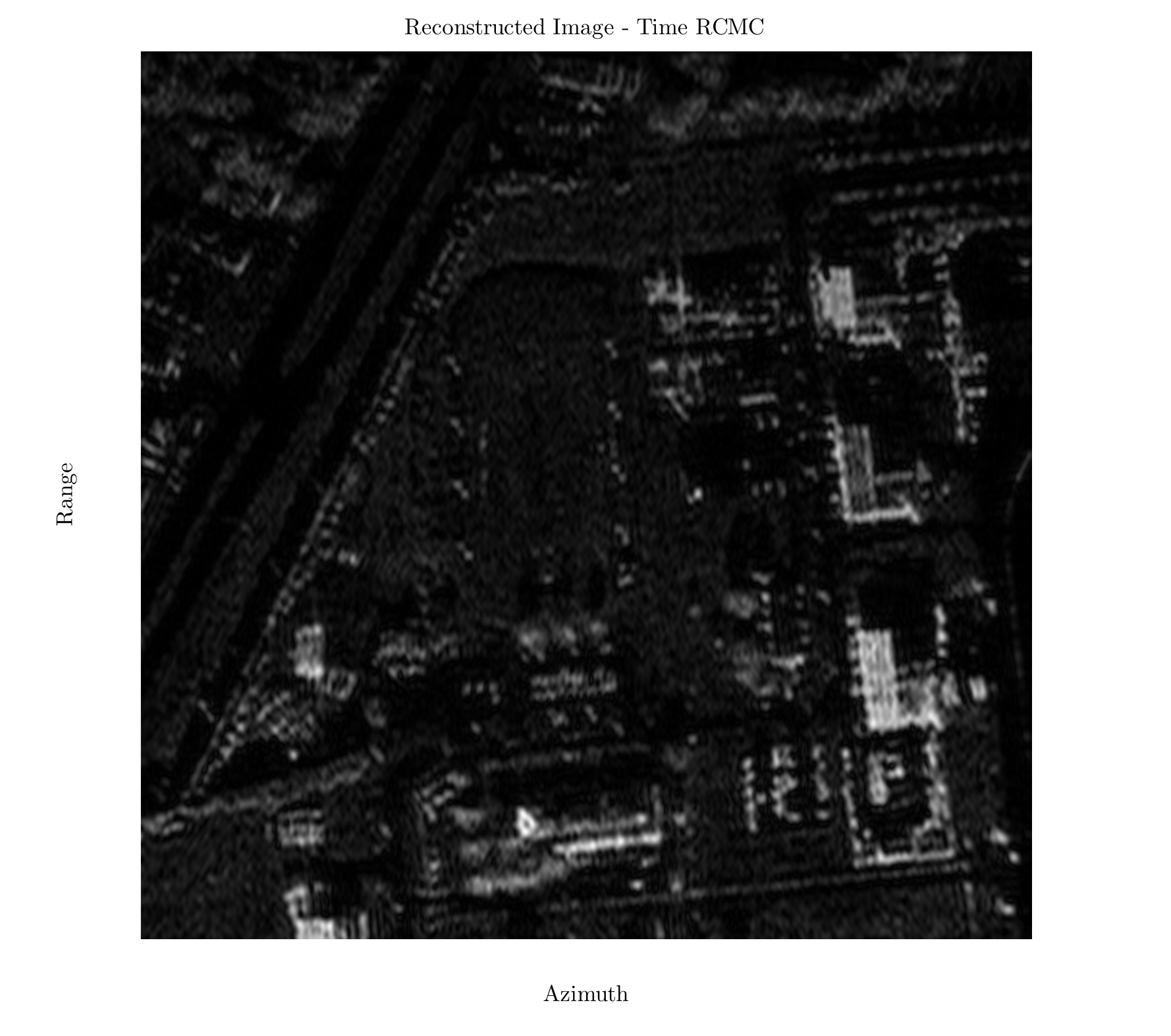} \\
(a) \\
\includegraphics[width=0.98\columnwidth]{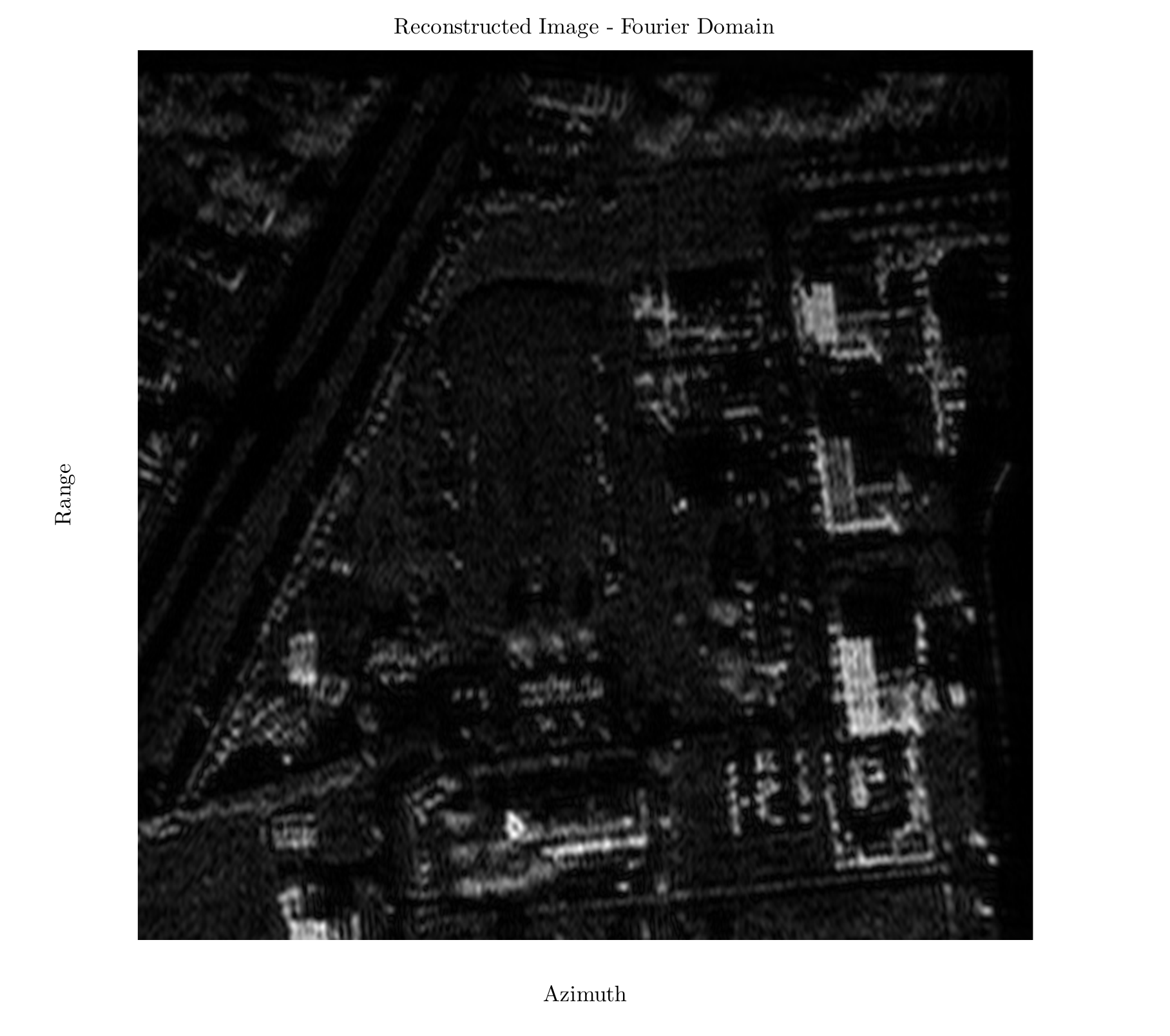} \\
 (b) \\
\end{tabular}
\caption{Comparison between two reconstructed images. (a) Conventional RDA using oversampled raw data. (b) Fourier RDA, $\left|\nu(k,l)\right| = 5$, with no oversampling.}
\label{fig:RCMCtimeVsFourierN}
\end{figure}

In addition, we examined the equivalence of both methods on SAR raw data which was simulated from a real SAR image as a reflectivity map, $\sigma(\br)$. To verify the selection of $\left|\nu(k,l)\right|$ we compare the resulting image of conventional RDA processing with Fourier domain RDA using a varied number of $\left|\nu(k,l)\right|$. We measured the similarity using a state of the art image quality assessment index call FSIM \cite{zhang2011fsim}. From Fig.~\ref{fig:FSIM} it is readily seen that the effect of considering more than 5 coefficients is negligible. The parameters of the system are described in Table \ref{tab:SARparameters}.
The ratio between the cardinality of the set $\beta
_m$ and the overall number of samples $N$, required by standard RDA rate $f_s$, is dictated by the oversampling factor, $\os$. Since $f_s=\os N$, the new rate leads to a reduction of $B/N=\frac{1}{2}$. Figure~\ref{fig:RCMCtimeVsFourierN}(a) shows the image follows conventional RDA processing while in Fig.~\ref{fig:RCMCtimeVsFourierN}(b) we use Fourier RDA processing with $\left|\nu(k,l)\right|=5$. As can be readily seen, the images look identical. These results verify that both signals and the resulting images are extremely similar.

\begin{figure}
\centering
 \includegraphics[width=0.7\linewidth]{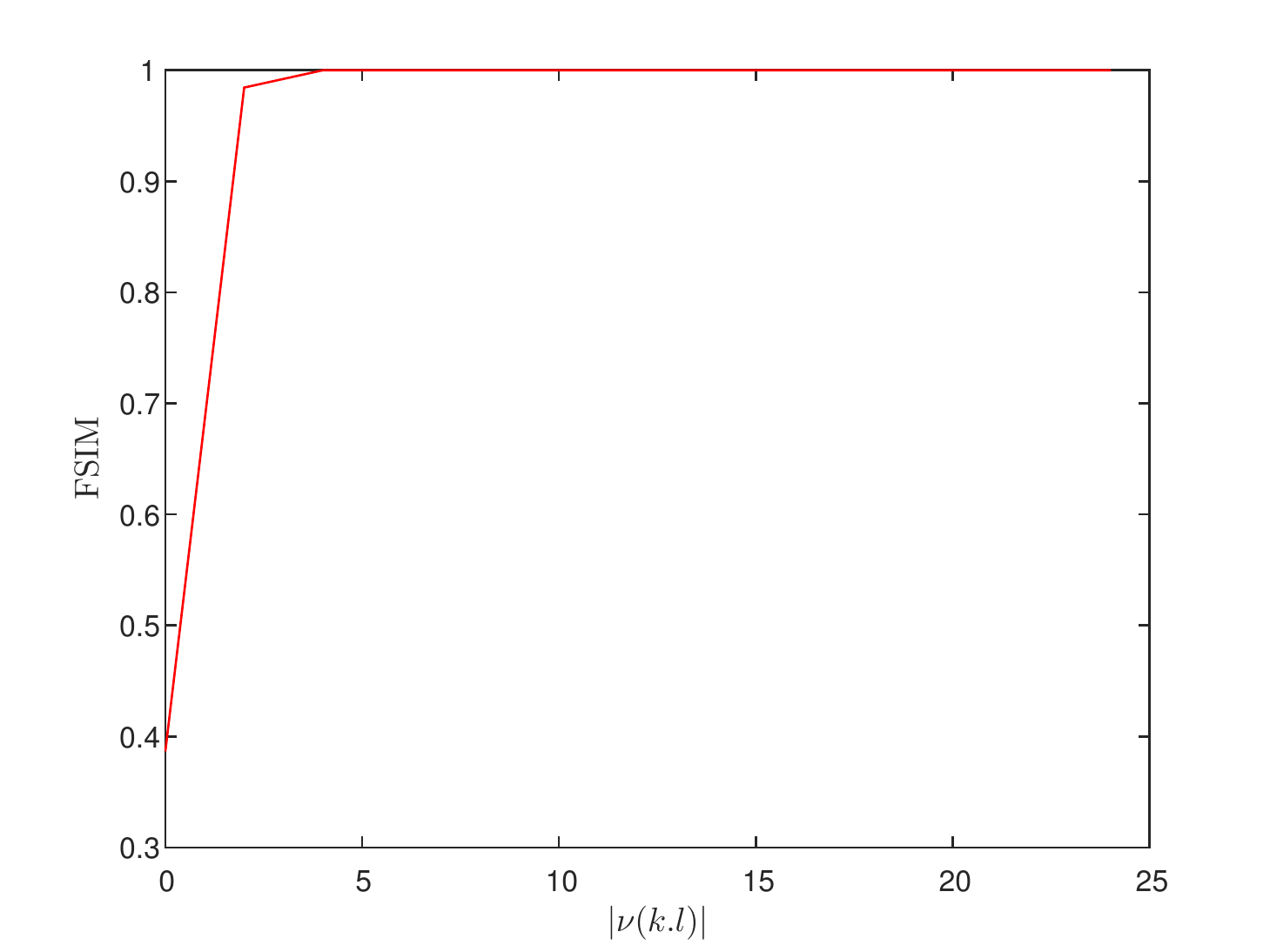}
\caption{Similarity measured by FSIM between conventional RDA and Fourier domain RDA for a varied number of Fourier coefficients which are considered before RCMC, $\left|\nu(k,l)\right|$.}
\label{fig:FSIM}
\end{figure}

To conclude this section, we presented a new algorithm, equivalent to RDA, that instead of time interpolation, can correct the migration of range cells in the Fourier domain. We exploited the effective bandwidth of SAR signals and bypassed oversampling, dictated by digital implementation of RCMC in time without any over sampling factor, assumption on the signal structure or the invariance of range blocks.

\section{Two-dimensional sub-Nyquist SAR}
\label{sec:2dsubNyquist}
We now demonstrate how Fourier domain RDA allows for sub-Nyquist sampling of the received signals, in both range and azimuth, when exploiting sparsity of SAR images. This two-dimensional reduction will enable perfect reconstruction using less pulses and fewer samples from each individual return.

\subsection{Sampling rate reduction via compressed sensing}
\label{subsec:rateRedCS}
Denote by $\tilde{\bbd} = \{\tilde{D}_m[l]\}\in\mathbb{C}^{B\times M}, 0\leq m<M, l\in\beta_k$, the Fourier coefficients of the range compressed signals in \eqref{eq:fourierDomainRangeCompression}.

Having $\tilde{\bbd}$, and using the processing stages in \eqref{eq:procRD}, \eqref{eq:azimuthCompression}, \eqref{eq:IFFTazimuth}, and \eqref{eq:scaledSigReconstruction}, the relationship between the image and the processed Fourier coefficients can be formulated as
\begin{equation}
	\tilde{\bbd} = \tilde{\bbq}\left( \bbf^s\left[\bbb\circ\left(\bbi\bbf\right)\right]\right)\bbf^{\ast},
	\label{eq:CSmainEq2D}
\end{equation}
where $\bbf^s=\{\frac{1}{B}e^{-j\frac{2\pi}{B}lk}\}\in \mathbb{C}^{B\times N}$ is a partial DFT matrix, $\bbb=\{e^{j\pi\frac{k^2}{K_a[n]}}\}\in\mathbb{C}^{N\times M}$ is the azimuth compression matrix from \eqref{eq:azimuthCompression}, $\bbf\in\mathbb{C}^{M\times M}$ is the DFT matrix, $\circ$ is the Hadamard product, $\bbi = \{I[n,m]\}\in\mathbb{C}^{N\times M}$ is the desired image and $\tilde{\bbq}\left(\cdot\right)$ is the inverse RCMC operator, which should satisfy 
\begin{equation}
	S_k[l] = \sum_{r=-\infty}^{\infty} C_k[r]\tilde{Q}_{k,l}[-r].
\label{eq:fourierInvRCMC}
\end{equation}
Under the low squint angle assumption (see Section~\ref{sec:subNyquistSAR}), the following proposition provides a simple expression for  $\tilde{Q}_{k,l}[r]$.
\textit{Proposition:}  
Suppose that $ak^2 \ll 1$ for every $0\leq k < M$. Then a good approximation for the inverse of the Fourier RCMC operator defined in \eqref{eq:fourierRCMC} is given by
\begin{equation}
\tilde{Q}_{k,l}[r]=(1+ak^2)e^{-j\pi\left(r+l(1+ak^2)\right)}\sinc\left(r+l(1+ak^2)\right).
\label{eq:Qgal}
\end{equation}
\textit{Proof:} To prove the result we need to show that for integer values of $n$ and $l$,
\begin{equation}
\label{eq:QQisDelta}
\sum_{r=-\infty}^{\infty}Q_{k,r}[-n] \tilde{Q}_{k,l}[-r]\approx\delta[n-l],
\end{equation}
where $\delta[n-l]$ is the Kronecker delta.

 For every $n,l\in\mathbb{Z}$ and $0\leq k< M$ we have
\begin{eqnarray}
\label{eq:QQisDelta}
\sum_{r=-\infty}^{\infty}Q_{k,r}[-n] \tilde{Q}_{k,l}[-r]= \sum_{r=-\infty}^{\infty}e^{-j\pi\left(-r-n+l(\frac{1}{1+ak^2}+1+ak^2)\right)}\\\nonumber\times\sinc\left(-n+\frac{l}{1+ak^2}\right)\sinc\left(-r+l(1+ak^2)\right).
\end{eqnarray}
Using the fact that $ak^2 \ll 1$ for every $0\leq k < M$, the last expression can be approximated by 
\begin{eqnarray}
\label{eq:QQisDelta}
 &\sum_{r=-\infty}^{\infty}&e^{-j\pi\left(-r-n+2l\right)}\sinc\left(-r+l\right)
\sinc\left(-n+l\right)  \\\nonumber
&=& \sinc\left(-n+l\right) e^{j\pi n}\sum_{r=-\infty}^{\infty}e^{-j\pi\left(-r+2l\right)}
\sinc\left(-r+l\right)  \\\nonumber
 &=& \sinc\left(-n+l\right)e^{j\pi n}e^{-j\pi l}.
\end{eqnarray}
Finally, the result follows from the fact that $\sinc(-n+l)=\delta[n-l]$ for integer values of $n$ and $l$.
\qedsymbol


When there exists some basis in which $\bbi$ is sparsely represented, \eqref{eq:CSmainEq2D} becomes a CS problem that can be solved using a smaller amount of rows and columns in $\tilde{\bbd}$. Using $\ell_1$ as a sparsity measure, the resulting optimization problem is:
\begin{equation}
	\min\left\| \bbpsi (\bbi)\right\|_1 \;\; \textrm{s.t.} \;\; \left\|\tilde{\bbd}_p-\tilde{\bbq}_{p}\left( \bbf^s\left[\bbb\circ\left(\bbi\bbf\right)\right]\right)\bbf^{\ast}_p\right\|^2_F < \epsilon
	\label{eq:optimizationProblemMain}
\end{equation}
where $\tilde{\bbd}_{p}$ is both a column and row under-sampled version of $\tilde{\bbd}$, $\tilde{\bbq_{p}}\left(\cdot\right)$ is the partial RCMC operator which considers only the subsampled Fourier coefficients, $\bbf^{\ast}_p$ is a column under-sampled version of $\bbf^{\ast}$, $\bbpsi$ is a sparsifying transform operator, $\left\|\cdot\right\|_F$ is the Frobenius norm and $\epsilon$ is an appropriate noise level which controls the fidelity of the reconstruction to the measured data. We denote the subsets of rows and columns of $\bbd_p$ by $\tilde{M} \subseteq \{1,2,\ldots ,M\}$ and $\tilde{\kappa} \subseteq \beta_m$, respectively.

There are various approaches to solve this optimization problem. In the field of SAR, most of the existing CS schemes stack the whole two-dimensional reflectivity map to a vector \cite{wei2010sparse,alonso2010novel,dong2014novel}. The vectorized form of \eqref{eq:optimizationProblemMain} is
\begin{equation}
	\min\left\|\bbpsi(\bx)\right\|_1 \;\; \textrm{s.t.} \;\;  \left\|\by-\bbm\bx\right\|_2 \leq\epsilon
	\label{eq:standardCS}
\end{equation}
where $\bx = \textrm{vec}\left(\bbi\right)$, $\by = \textrm{vec}\left(\bbc\right)$ and $\bbm = \bar{\bbf}^*_p\bar{\bbq}_p \bar{\bbf^s}\bar{\bbb}\bar{\bbf}$ with
$\bar{\bbf}= \bbf^T\otimes \tilde{\bbi}$,
$\bar{\bbb}=\diag\{\textrm{vec}(\bbb)\}$, $\bar{\bbf^s}=\tilde{\bbi}\otimes\bbf^s$, $\bar{\bbq}_p=\diag\{\bbq^{(k)}\}$, $\bar{\bbf}^*_p= {\bbf^\ast_p}^T\otimes \tilde{\bbi}$, where $\otimes$ is Kronecker product and $\tilde{\bbi}$ is the identity matrix.
A variety of CS techniques can then be employed to solve \eqref{eq:standardCS}, such as interior point methods \cite{boyd2004convex} and alternating direction method of multipliers (ADMM) \cite{boyd2011distributed}, \cite{afonso2010fast}. Fast iterative shrinkage-thresholding algorithms such as FISTA \cite{beck2009fast,palomar2010convex} or its monotonic version MFISTA \cite{tan2014smoothing} are more favorable in dealing with large dimensional data since they do not require structure. Due to the long reconstruction time and large memory requirements, it is difficult to reconstruct a moderate-size scene using CS and vectorization in practice.

Instead, we next show how to solve \eqref{eq:optimizationProblemMain} by extending FISTA \cite{beck2009fast} to support two-dimensional matrix recovery which fits the SAR problem without the use of vectorization. We apply the same technique as in \cite{wimalajeewa2013recovery}. The proposed algorithm is coined SAR FISTA.

In general, FISTA is aimed at minimizing an error function, which in our case equals
\begin{equation}
	\bbg\left(\bbi\right) = \left\|\tilde{\bbd}_p-\tilde{\bbq}\left( \bbf^s\left[\bbb\circ\left(\bbi\bbf\right)\right]\right)\bbf^{\ast}_p\right\|^2_F.
	\label{eq:matrixNormPRF}
\end{equation}
It relies on soft thresholding and gradient decent. The soft operator for a matrix $\bbx$ is defined via
\begin{equation}
	\textrm{soft}\left(\bbx,\alpha\right) = \frac{\bbx_{ij}}{\left|\bbx_{ij}\right|}\left(\left|\bbx_{ij}\right|-\alpha\right)_+.
	\label{eq:soft}
\end{equation}
The Lipshitz constant of $\bbg\left(\bbi\right)$, $L_f$, controls the gradient decent step of the error function, which is given by
\begin{equation}
\nabla\bbg\left(\bbi\right) = 2\left\{\bbb\circ\left[{\bbf^s}^H\tilde{\bbq}_p^H\left(\bbe\bbf_p^{T}\right)\right]\right\}\bbf^H,
	\label{eq:gradientPRF}
\end{equation}
where $H$ is the adjoint operator and
\begin{equation}
\bbe = \tilde{\bbd}_p-\tilde{\bbq}_p\left( \bbf^s\left[\bbb\circ\left(\bbi\bbf\right)\right]\right)\bbf^{\ast}_p.
	\label{eq:innerGradientPRF}
\end{equation}

Since reconstruction is performed using $\tilde{\bbd}_p$ as the measurements, range compression should be performed as a preprocessing stage. For a given subsampled data, $\bbd_p$, our steps for reconstruction using SAR FISTA are summarized in Algorithm~\ref{alg:SARPRFFISTA}. The runtime of Algorithm~\ref{alg:SARPRFFISTA} is dictated by step 4, which considers the derivative of the RCMC operator. Since the gradient decent step is repeated iteratively, this operation constitutes the bottle neck of Algorithm~\ref{alg:SARPRFFISTA}. In Section~\ref{sec:subNyquistSAR} we show how to reduce runtime in the case in which only range subsampling is required.

\begin{algorithm}
\caption{SAR FISTA reconstruction for two-dimensional sub-Nyquist sampling}
\label{alg:SARPRFFISTA}

\textbf{Input:} SAR raw data xamples $\bbd_p=\left\{D_m[l]\right\}_{m\in \tilde{M}}^{l\in\tilde{\kappa}}$, measurement matrices $\bbf^s_p$, $\bbb$, $\bbf$\\
\textbf{Output:} estimate for sparse coefficients of SAR image, $\hat{\bbx}$, such that $\bbi = \bbpsi^{-1}(\hat{\bbx})$\\
\begin{algorithmic}[1]
\STATE \textbf{Initialization:} $\tilde{\bbd}_p = \left\{\tilde{D}_m[l]\right\}_{m\in \tilde{M}}^{l\in\kappa}\leftarrow\bbd_p$ via \eqref{eq:fourierDomainRangeCompression} \\
			 \textbf{Initialize:} $\bbx^0=\bf{0}$, $\bbx^1=\bf{0}$, $t_0 = 1$, $t_1=1$, $k=1$\\
			 $\lambda_1,\beta\in\left(0,1\right)$, $\bar{\lambda}>0$
\WHILE {not converged}
\STATE $\bbz^k = \bbx^k + \frac{t_{k-1}-1}{t_k}\left(\bbx^k-\bbx^{k-1}\right)$
\STATE $\bbu^k = \bbz^k -\frac{1}{L_f}\nabla\bbg\left( \bbpsi^{-1}(\hat{\bbx})\right)$, via \eqref{eq:gradientPRF}
\STATE $\bbx^{k+1} = \textrm{soft}\left(\bbu^k,\frac{\lambda_k}{L_f}\right)$
\STATE $t_{k+1} = \frac{1+\sqrt{4t_k^2+1}}{2}$
\STATE $\lambda_{k+1} = \max\left(\beta\lambda_k,\bar{\lambda}\right)$
\STATE $k = k+1$
\ENDWHILE
\end{algorithmic}
$\hat{\bbx} = \bbx$
\end{algorithm}

Algorithm~\ref{alg:SARPRFFISTA} enables reconstruction of a SAR image, using less columns and rows of the raw data matrix, which correspond to the emission of pulses and Fourier samples, respectively. We next explain how the reduction in Fourier samples is equivalent to the reduction of time domain samples of the individual returns.

\subsection{Analog-to-Digital rate reduction}
\label{subsec:A2D}
In the previous sections, we assumed that a subset of Fourier coefficients are given. However, in order to construct a real sub-Nyquist sampling system, these coefficients should be derived from low-rate time domain samples generated from a low rate ADC. We next explain how the required Fourier coefficients can be extracted from the raw data samples in time.

Similarly to \cite{wagner2012compressed}, we use the Xampling mechanism. The Xampling philosophy ties together sub-Nyquist sampling based on analog preprocessing with techniques of CS for recovery. However, these approaches typically require sophisticated sampling schemes, which acquire generalized measurements of the analog signals \cite{eldar2015sampling,gedalyahu2011multichannel}. The authors in \cite{baransky2014sub} presented a concrete analog-to-digital conversion scheme and a recovery algorithm for sampling radar signals at sub-Nyquist rates. The analog input is split into channels, where in each channel it is mixed with the selected harmonic signal, integrated over the PRI duration, and then sampled. The matching Fourier coefficients are then created digitally.

Using Xampling, the next question is which frequencies should be selected, considering practical limitations. In CS, the natural selection is to choose the coefficients randomly. Unfortunately, this sampling strategy is not practical in hardware. Some guidelines for choosing the frequencies, were suggested in \cite{stoica2012sparse} in order to solve the trade-off between noise robustness, which is increased by highly distributed frequency samples and practical hardware implementation. This trade-off is also between high resolution, which requires a wide aperture, and avoiding ambiguities, which calls for close frequencies. Coping with the practical limitation, similarly to \cite{baransky2014sub}, a multiple bandpass sampling approach was chosen, where four groups of consecutive coefficients are selected. The board can be seen in Fig.~\ref{fig:hardware}(a).

\begin{figure}
\centering
\begin{tabular}{ccc}
\includegraphics[width=0.32\columnwidth]{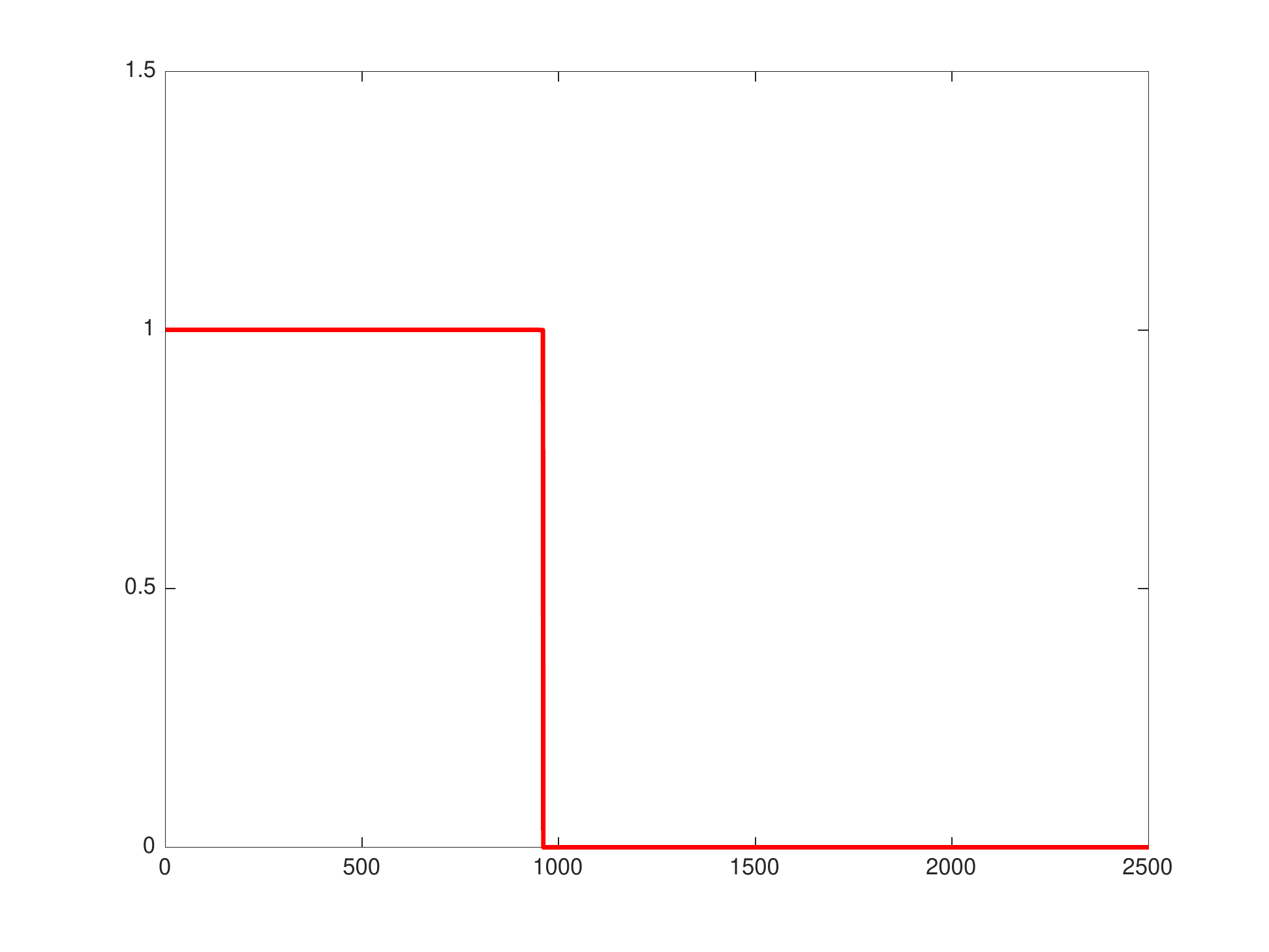} &
\includegraphics[width=0.32\columnwidth]{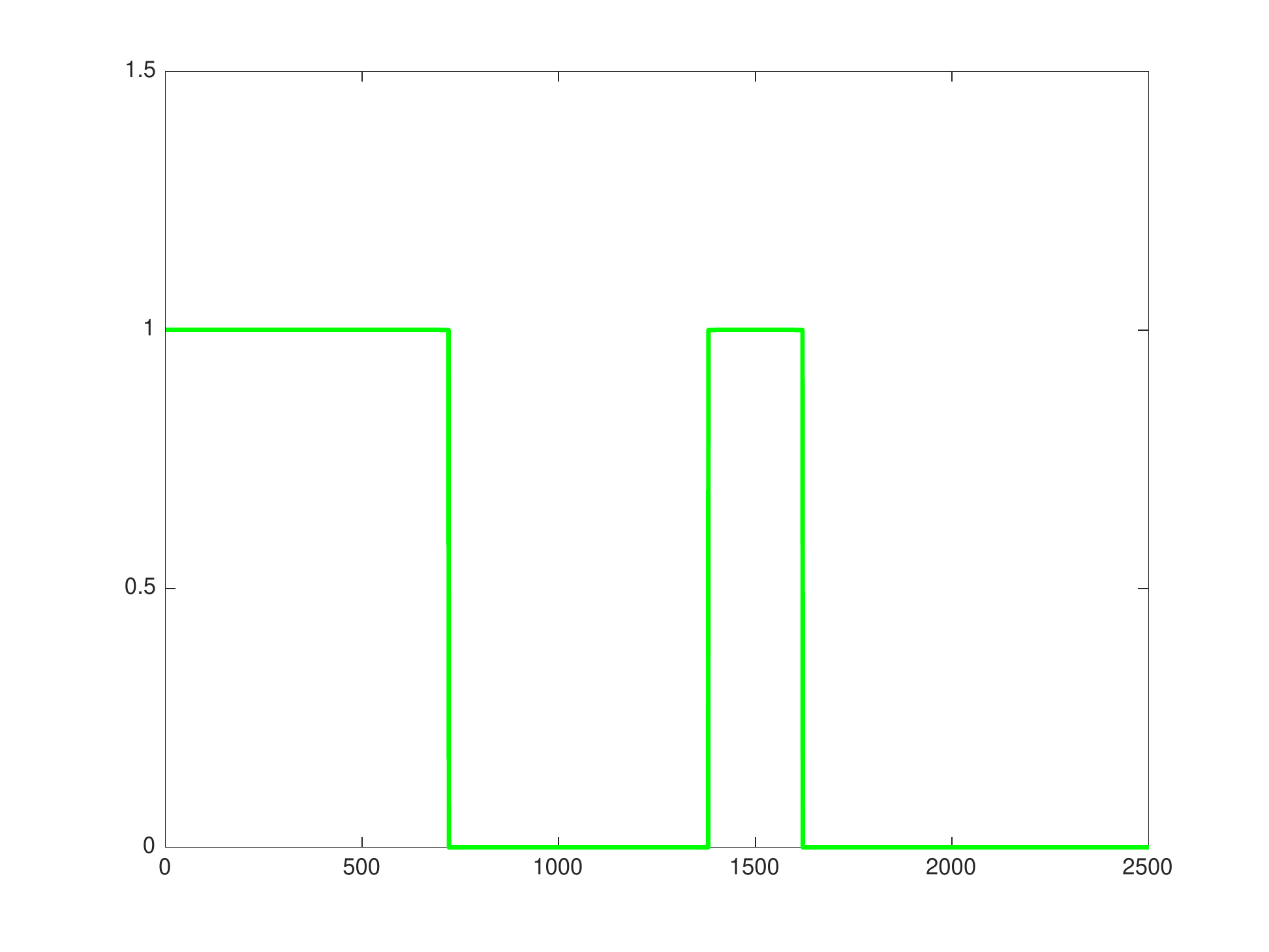} &
\includegraphics[width=0.32\columnwidth]{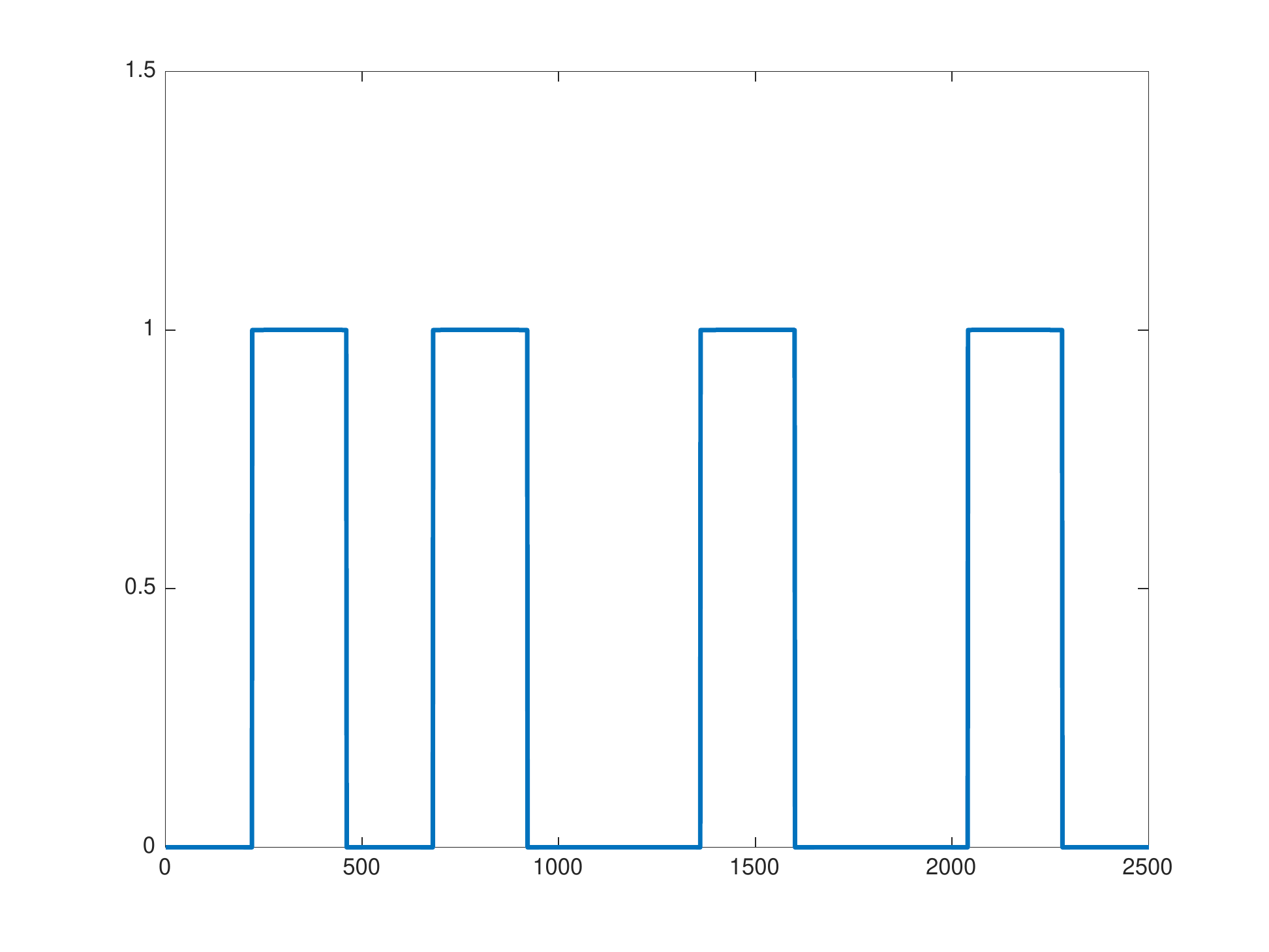} \\
(a)&(b)&(c)
\end{tabular}
\caption{Frequency bands selection. (a) Low pass (``PDF \#1"). (b) A combination of low pass and a band pass (``PDF \#2"). (c) Random selection (``PDF \#3").}
\label{fig:PDFs}
\end{figure}

We next examine three different practical sampling scenarios, each consisting of 4 frequency bands. We transmit a linear chirp into a one-dimensional synthesized scene, where $30\%$ of the scene samples are zero in the time domain, as well as in the wavelet domain, under the Daubechies 4 basis. The first frequency samples selection (marked as ``PDF \#1") includes the lower half of the frequency samples, the second selection (``PDF \#2") includes a lowpass and a narrower bandpass and the third selection (``PDF \#3") includes randomly selected 4 bands of frequency samples. An illustration of the selected frequency bands for each of the scenarios is depicted in Fig.~\ref{fig:PDFs}.

The reconstruction of the scene is performed using three methods: direct reconstruction using matched filtering with an appropriate subsampled chirp signal, reconstruction with FISTA using Daubechies-4 wavelets as the sparsifying transform, and reconstruction with FISTA using an identity transform.

\begin{figure}
\centering
\includegraphics[scale=0.7]{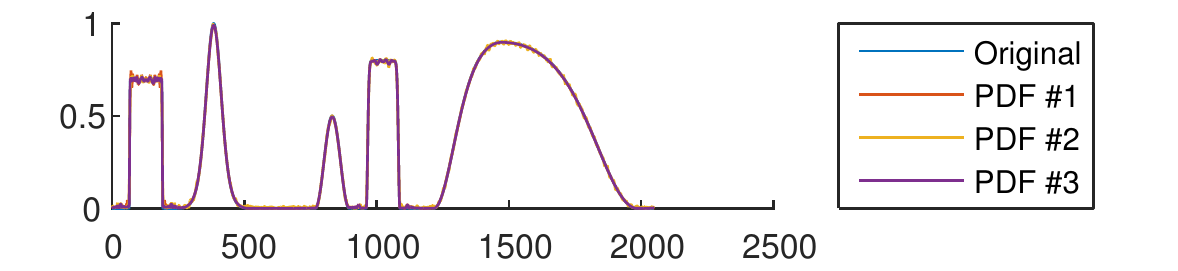} \\
\includegraphics[scale=0.7]{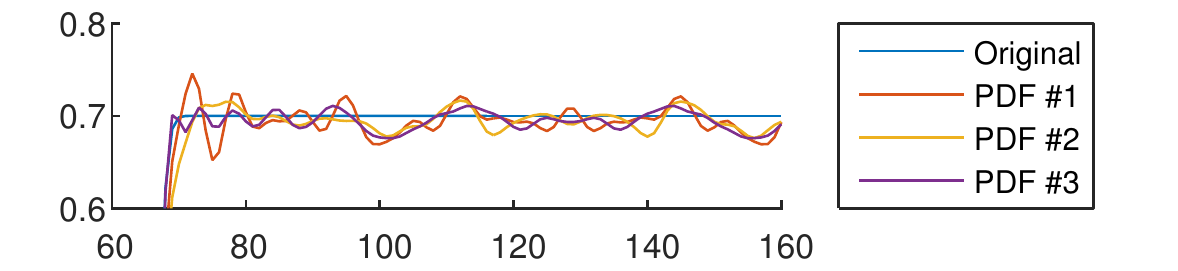} \\ (a)\\
\includegraphics[scale=0.7]{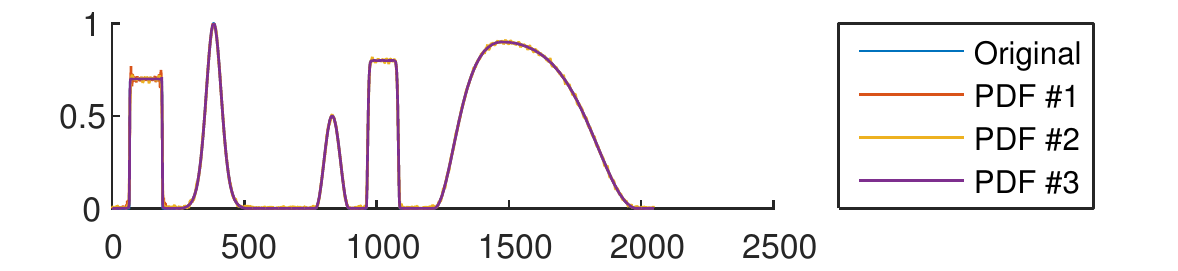} \\
\includegraphics[scale=0.7]{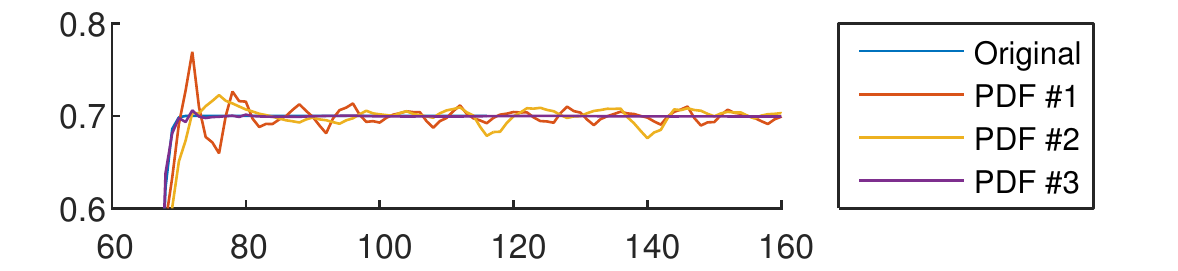} \\ (b)
\includegraphics[scale=0.7]{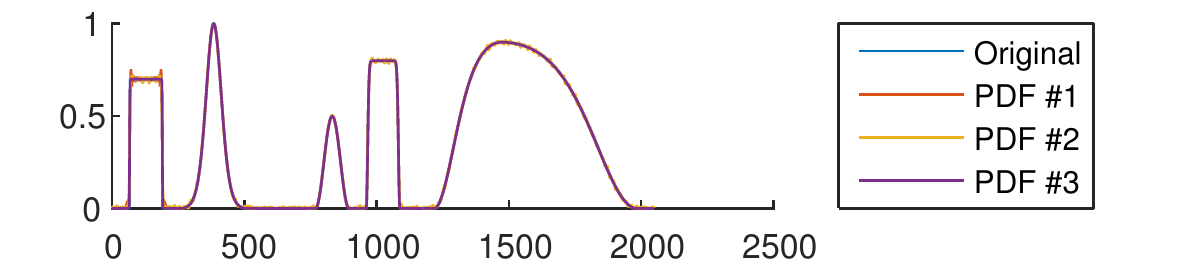} \\
\includegraphics[scale=0.7]{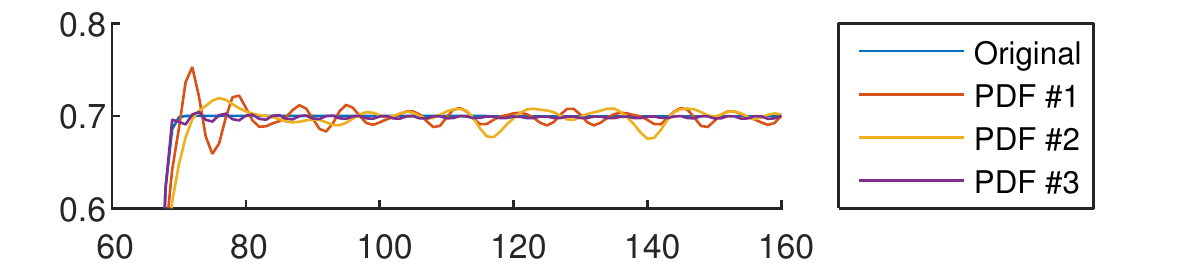} \\ (c)
\caption{One-dimensional signal reconstruction based on different subsampling strategies (paired with zoomed part). (a) Direct reconstruction using matched filtering. (b) Reconstruction with FISTA under the wavelet transform. (c) Reconstruction with FISTA under an identity transform.}
\label{fig:1D_CS_Zoom}
\end{figure}

As seen in Fig.~\ref{fig:1D_CS_Zoom}, the reconstruction quality using FISTA surpasses direct reconstruction both qualitatively (the signals recovered with FISTA show less ripple), as well as numerically: with PDF \#3, the error norm for direct reconstruction is around $0.228$ while it is $0.084$ for reconstruction with FISTA under Wavelets, and  $0.070$ for reconstruction with FISTA under an identity transform. The combination of CS reconstruction (with an arbitrary sparsifying $\Psi$), along with random bands selection, which best copes with the mentioned trade-offs, provides the best performance for sub-Nyquist in range. The randomness encourages dynamic changes which are not limited to certain bands. We will use this property in Section~\ref{sec:transmitter}, when we present the cognitive SAR concept.

\subsection{Performance Improvement}
\label{sec:subNyquistSAR}
As mentioned in Section~\ref{subsec:rateRedCS}, the heaviest part in terms of runtime in Algorithm~\ref{alg:SARPRFFISTA} is the gradient decent step which is performed every iteration. Following  \cite{aberman2016range}, in the case that only range subsampling is required and under certain assumptions, we can simplify the algorithm. In particular, we next present a method that exploits the structure of the RCMC operator in \eqref{eq:fourierRCMC}, in order to take this operator out of the gradient step and apply it only once.

Denote by $\bbc = \{C_k[l]\}_{0\leq k<M}^{l\in\beta_k}\in\mathbb{C}^{B\times M}$ the partial Fourier coefficients matrix of the corrected signals and by $\bbq(\cdot)$ the RCMC operator which is defined via \eqref{eq:fourierRCMC}. Since the DFT is a unitary matrix, right multiplying by $\bbf$ and applying $\bbq(\cdot)$ on \eqref{eq:CSmainEq2D} leads to
\begin{equation}
	\bbc = \bbf^s\left[\bbb\circ\left(\bbi\bbf\right)\right],
	\label{eq:CSmainEq}
\end{equation}
where $\bbc = \bbq(\tilde{\bbd}\bbf)$. Repeating the same steps as in Section~\ref{subsec:rateRedCS}, the optimization problem in \eqref{eq:optimizationProblemMain} becomes
\begin{equation}
	\min\left\|\bbpsi(\bbi)\right\|_1 \;\; \textrm{s.t.} \;\; \left\|\bbc_p-\bbf^s_p\left[\bbb\circ\left(\bbi\bbf\right)\right]\right\|^2_F < \epsilon,
	\label{eq:optimizationProblemMatrix}
\end{equation}
where $\bbc_p$ and $\bbf^s_p$ are row undersampled versions of $\bbc$ and $\bbf^s$.

In this case as well, we can reconstruct the image using FISTA, where the gradient of the error function is
\begin{equation}
	\nabla\bbf\left(\bbi\right) = 2\left\{\bbb\circ\left[{\bbf^s}^H\left(\bbf^s\left(\bbb\circ\left(\bbi\bbf\right)\right)-\bbc\right)\right]\right\}\bbf^H.
	\label{eq:matrixGradient}
\end{equation}
It can be seen that \eqref{eq:fourierInvRCMC} is not part of the gradient step in \eqref{eq:matrixGradient}. The RCMC operator is performed only in the preprocessing stage to create $\bbc$. However, due to the fact that subsampling is not performed on the raw data itself, we next have to figure out how many Xamples, $D_m[l]$, should be considered in order to extract $\bbc_p$. To answer this question, we examine $\kappa \subset\beta_m$, a subset of $D_m[l]$.

Due to the decay property of \eqref{eq:Qs}, the relationship in \eqref{eq:fourierRCMC} implies that calculation of a specific Fourier coefficient $C_k[l]$, requires only $\left|\nu(k,l)\right|$ coefficients of $\left\{S_k[l]\right\}_l$. The decay rate of \eqref{eq:Qs} and thus the cardinality of $\nu(k,l)$ is dictated by the behavior of the sinc function and is independent of $k$ or $l$. We denote the cardinality of $\nu(k,l)$ by $L$. Thus, for a given Doppler frequency $k$, in order to compute an arbitrary set of B coefficients from $\left\{C_k[l]\right\}_l$, only $B+L$ coefficients $\left\{S_k[l]\right\}_l$ are needed.

Considering the azimuth DFT in \eqref{eq:azimuthDFTonFourier} it is easy to see that in order to extract a specific coefficient $S_k[l]$ we need the entire set of $\left\{D_m[l]\right\}_m$. Therefore, to evaluate an individual coefficient, $C_k[l]$, the  indices of the coefficients which should be xampled from each individual signal are $\nu(k,l)$. Generalizing the concept for the entire matrix, in order to extract $\bbc_p=\left\{C_k[l]\right\}_{0\leq k<M}^{l\in\kappa}$, where $\kappa\subset\beta_k$, only $\bbd_p = \left\{D_m[l]\right\}_{0\leq m<M}^{l\in\tilde{\kappa}}$ should be xampled, where $\tilde{\kappa}\subset\beta_m$. We next show that $\kappa$ and $\tilde{\kappa}$ are of the same order of magnitude, which means that the preprocessing stages do not influence the number of required Xamples.

From the low squint angle assumption we have that when the squint angle is low, the range cell migration is relatively small. We may therefore assume that $ak^2 \ll 1$ for every $0\leq k<M$. To justify this assumption note that for the stripmap mode $ak^2 = \frac{1}{8}\left(\frac{\lambda}{vT}\right)^2\left(\frac{k}{M}\right)^2 < \frac{1}{8}\left(\frac{\lambda}{vT}\right)^2$ for every $k$. In the SEASAT-A satellite \cite{cumming1979digital}, $\lambda = 0.235$ [m], $v=7000$ [m/s], $T=0.6$ [msec], which yields $\frac{1}{8}\left(\frac{\lambda}{vT}\right)^2 = 3.8\times 10^{-4}$, justifying the approximation. This assumption means that the most dominant coefficient, $Q_{k,l}[-n]$, is $n_{k,l} \approx l$, a fact which implies that $\nu(k,l)$ is approximately independent of $k$, and leads to the approximation that the azimuth DFT operation does not influence the number of required xamples,
\begin{equation}
	\left|\kappa_m\right| =\left| \bigcup_{k,l}\nu(k,l)\right| \approx B+L.
	\label{eq:indices}
\end{equation}
As was shown in Section~\ref{subsec:SimulationAndValidation}, we selected $L$ to be 5. This means that when $B \gg L$, the preprocessing stages do not drastically enlarge the number of required Xamples.

\begin{algorithm}
\caption{SAR FISTA for sub-Nyquist sampling in range}
\label{alg:SARFISTA}

\textbf{Input:} Xamples $\bbd_p=\left\{D_m[l]\right\}_{0\leq m<M}^{l\in\tilde{\kappa}}$, measurement matrices $\bbf^s_p$, $\bbb$ and $\bbf$\\
\textbf{Output:} estimate for sparse coefficients of SAR image, $\hat{\bbx}$, such that $\bbi = \bbpsi^{-1}(\hat{\bbx})$\\
\begin{algorithmic}[1]
\STATE \textbf{Initialization:} $\bbc_p = \left\{C_k[l]\right\}_{0\leq k<M}^{l\in\kappa}\leftarrow\bbd_p$ via \eqref{eq:fourierRCMC}, \eqref{eq:fourierDomainRangeCompression} and \eqref{eq:azimuthDFTonFourier}\\
			 \textbf{Initialize:} $\bbx^0=\bf{0}$, $\bbx^1=\bf{0}$, $t_0 = 1$, $t_1=1$, $k=1$\\
			 $\lambda_1,\beta\in\left(0,1\right)$, $\bar{\lambda}>0$
\WHILE {not converged}
\STATE $\bbz^k = \bbx^k + \frac{t_{k-1}-1}{t_k}\left(\bbx^k-\bbx^{k-1}\right)$
\STATE $\bbu^k = \bbz^k -\frac{1}{L_f}\nabla\bbf\left( \bbpsi^{-1}(\hat{\bbx})\right)$, via \eqref{eq:matrixGradient}
\STATE $\bbx^{k+1} = \textrm{soft}\left(\bbu^k,\frac{\lambda_k}{L_f}\right)$, via \eqref{eq:soft}
\STATE $t_{k+1} = \frac{1+\sqrt{4t_k^2+1}}{2}$
\STATE $\lambda_{k+1} = \max\left(\beta\lambda_k,\bar{\lambda}\right)$
\STATE $k = k+1$
\ENDWHILE
\end{algorithmic}
$\hat{\bbx} = \bbx$
\end{algorithm}

Algorithm~\ref{alg:SARFISTA} describes the modified version of FISTA which supports the structure of \eqref{eq:optimizationProblemMatrix}. Unlike Algorithm~\ref{alg:SARPRFFISTA}, it uses the expensive three-dimensional operator in \eqref{eq:fourierRCMC} only once at the initialization
stage while calculating $\bbc_p$. For that reason, if we only
want to subsample in range, then Algorithm~\ref{alg:SARFISTA} is preferred.

\section{Exploiting Gaps in Time and Frequency}
\label{sec:gaps}
In Section~\ref{sec:2dsubNyquist} we presented a sub-Nyquist framework which allows two-dimensional subsampling along with reconstruction. As a result of the missing pulses and the reduced number of Fourier coefficients, time gaps (during CPI) and frequency holes (within the received signal's spectrum) exist in our system. In this section we explain how to exploit these gaps in each dimension.

\subsection{Reduced Time-on-Scene}
\label{subsec:reducedTimeonScene}

Algorithm~\ref{alg:SARPRFFISTA} enables reconstruction of a sparse scene with a number pulses which is less than the Nyquist requirement. This sub-Nyquist sampling in the azimuth direction is, in practice, a non-uniform transmission which results in time gaps within the CPI where no echoes are recorded. This can be interpreted as a \textit{reduced time-on-scene} concept, which stands for the reduction of time that the radar beam needs to steer at the scatters within the scene. Similarly to \cite{cohen2016RToT}, which uses the same concept for radar signals, we propose to exploit these time gaps, for sending pulses to other scenes. This allows to capture several different regions within the same CPI and therefore using the same size of memory to form several images instead of one. This memory reduction has significant meaning in orbital missions which are limited by on-board memory and downlink throughput. The processing of each image is performed separately, since every scene is processed with its own partial Fourier IDFT matrix, $\bbf_p^{\ast}$ in \eqref{eq:optimizationProblemMain}, with the indices of the relevant pulses.

\begin{figure}
\centering
\includegraphics[width=\columnwidth]{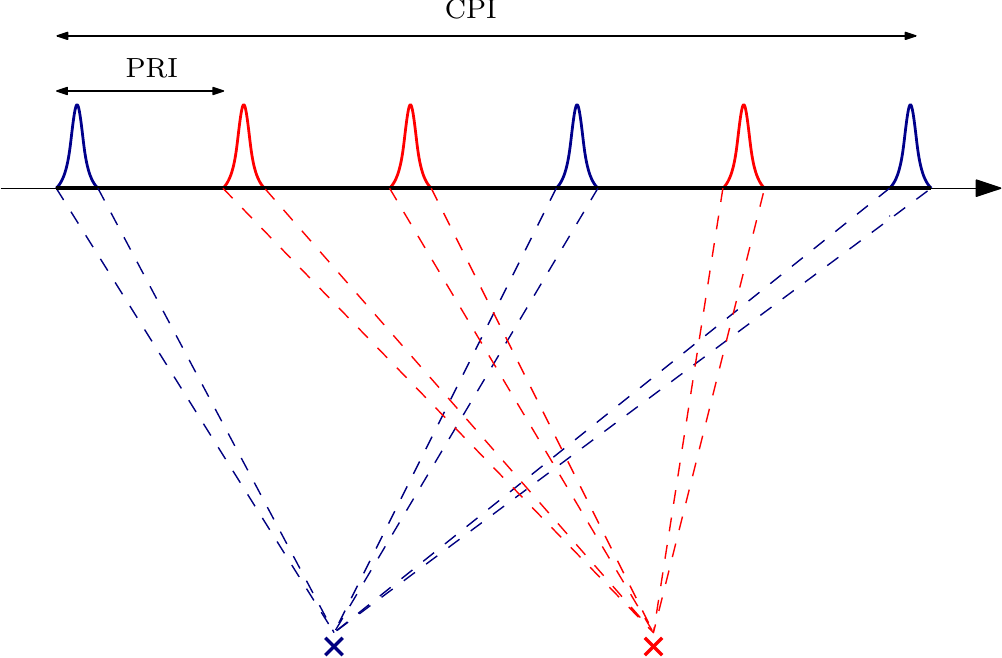}
\caption{Reduced time-on-scene. The transmitted pulses are non-uniformly sub-sampled. The complementary pulses are exploited to capture another scene. }
\label{fig:reducedPRF}
\end{figure}

Although these time gaps are on the order of milliseconds, phased array and electronic beam-steering techniques can aim the beam to different directions within those time periods, by controlling the phased array parameters \cite{younis2003digital}.
In the simulations, we show that two different scenes can be captured during a single CPI. Figure~\ref{fig:reducedPRF} depicts the reduced time on scene and time gaps exploitation concept.

In the next section we will demonstrate via simulations, how sub-Nyquist in azimuth is exploited in order to capture a wider area within the same CPI.

\subsection{Frequency Adaptive Transmitter}
\label{sec:transmitter}
We next show how to exploit our sub-Nyquist range abilities to allow for dynamic adaptation of both the transmitted and
received signal spectrum, paving the way to cognitive SAR. In particular, similarly to \cite{cohen2016CR} we modify the transmitter of the radar prototype presented in \cite{baransky2014sub} to adapt it to CR. Combining the transmission of a few narrow bands and using the reconstruction method described in Section~\ref{sec:subNyquistSAR}, we propose to enable dynamic spectrum changes of the transmitted SAR waveform. This will not affect any aspect of our sub-Nyquist processing since the received signal is preserved in the bands of interest. Let $\tilde{H}(\omega,t)$ be the CTFT of the new transmitted radar pulse,
\begin{equation}
\label{eq:bands}
\tilde{H}(\omega,t) =
     \begin{cases}
       H(\omega) &  \omega \in \mathcal{N}_b(t)\\
       0 & \text{otherwise},
     \end{cases}
\end{equation}
where
\begin{equation}
\mathcal{N}_b(t) = \bigcup_{1\leq i \leq N} [f_x^i(t) - B_x(t)/2, f_x^i(t)+B_x(t)/2] \nonumber
\end{equation}
 is the dynamic support of filtered $N$ bands, $B^i_x(t)$ and $f^i_x(t)$  are the bandwidth and the carrier frequency of $i$th band at time $t$, respectively. Obviously, the computation of the relevant Fourier coefficients $D_m[l]$ will not change.

To comply with CR requirements, the band parameters $B^i_x(t)$ and $f^i_x(t)$ vary with time allowing dynamic adaptation to the environment. Moreover, in Section~\ref{subsec:A2D} it was shown that the best practical sampling strategy consists of a random selection of a group of bands. This strategy enhances the ability to dynamically adapt the bands to vacant frequencies and best fits our cognitive system.

This approach leads to two main advantages. First, since we only use the received bands to transmit, the entire power is concentrated in them. Therefore, the SNR in the sampled bands is improved. Second, this technique allows for a dynamic form of the transmitted signal spectrum, where only a small portion of the whole bandwidth is used at each transmission. In the following section we demonstrate how sub-Nyquist in range is exploited in order to adapt cognition while improving SNR.

\section{Software and Hardware Simulations}
\label{sec:SimulationResults}
In this section, we examine the performance of Fourier domain RDA sub-Nyquist sampling for both the range and azimuth axes using simulated and real SAR data. We compare our methods to conventional RDA with full Nyquist samples. In addition, we present our hardware prototype and demonstrate the advantages of the proposed cognitive SAR in terms of SNR.

\subsection{Simulated Data}
In order to examine our methods, we generated SAR raw data of two different scenes: a spatially sparse scene and an image which is sparse under the Daubechies-4 wavelet basis. The data was generated from real SAR images, using the model in \eqref{eq:received} where the reflectivity map was taken as the original image, namely, each pixel in the image is treated as a point reflector, $\sigma(\br)=I(\br)$.

\begin{figure}
\centering
\begin{tabular}{c}
\includegraphics[width=\columnwidth]{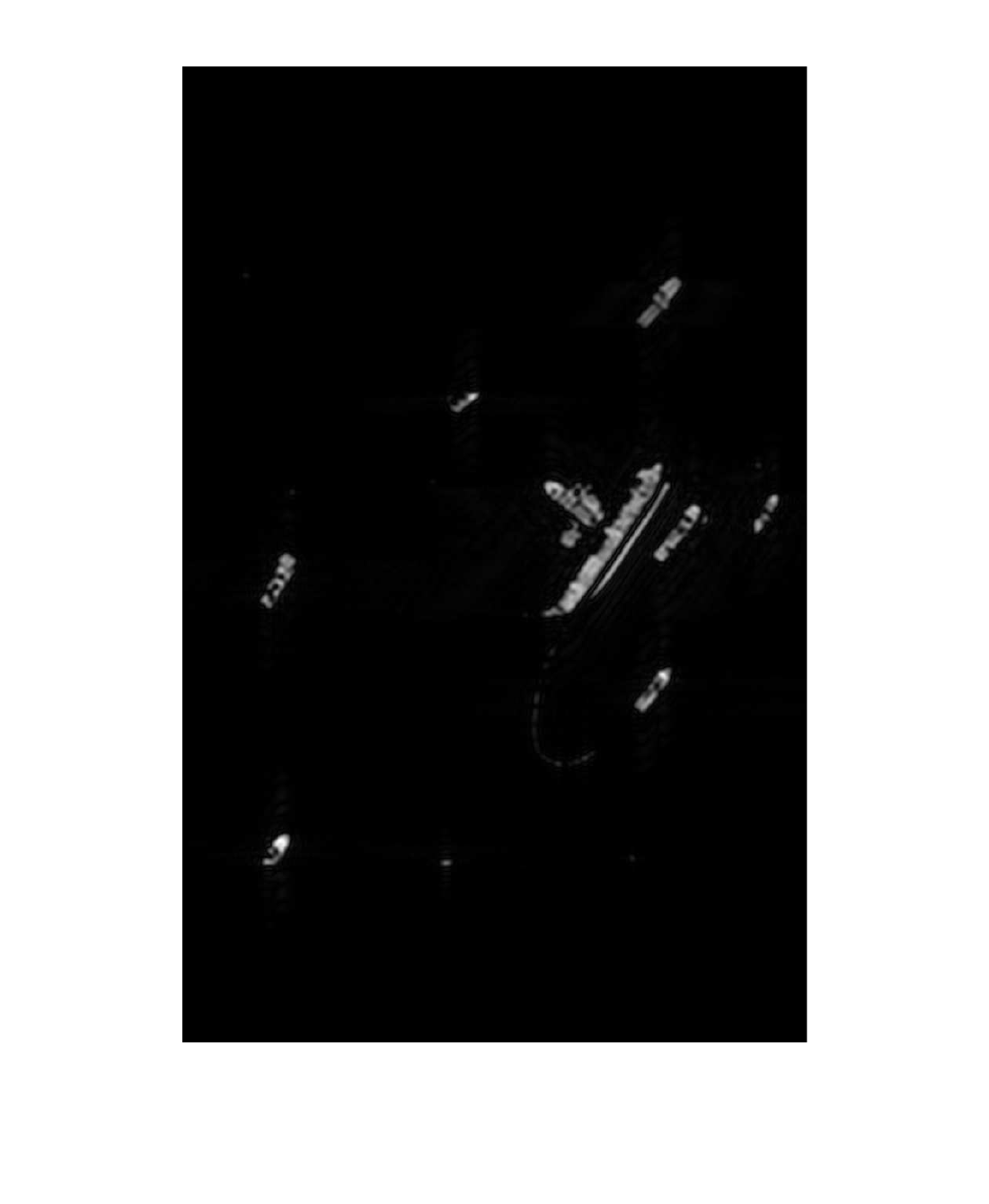} \\
(a) \\
\includegraphics[width=\columnwidth]{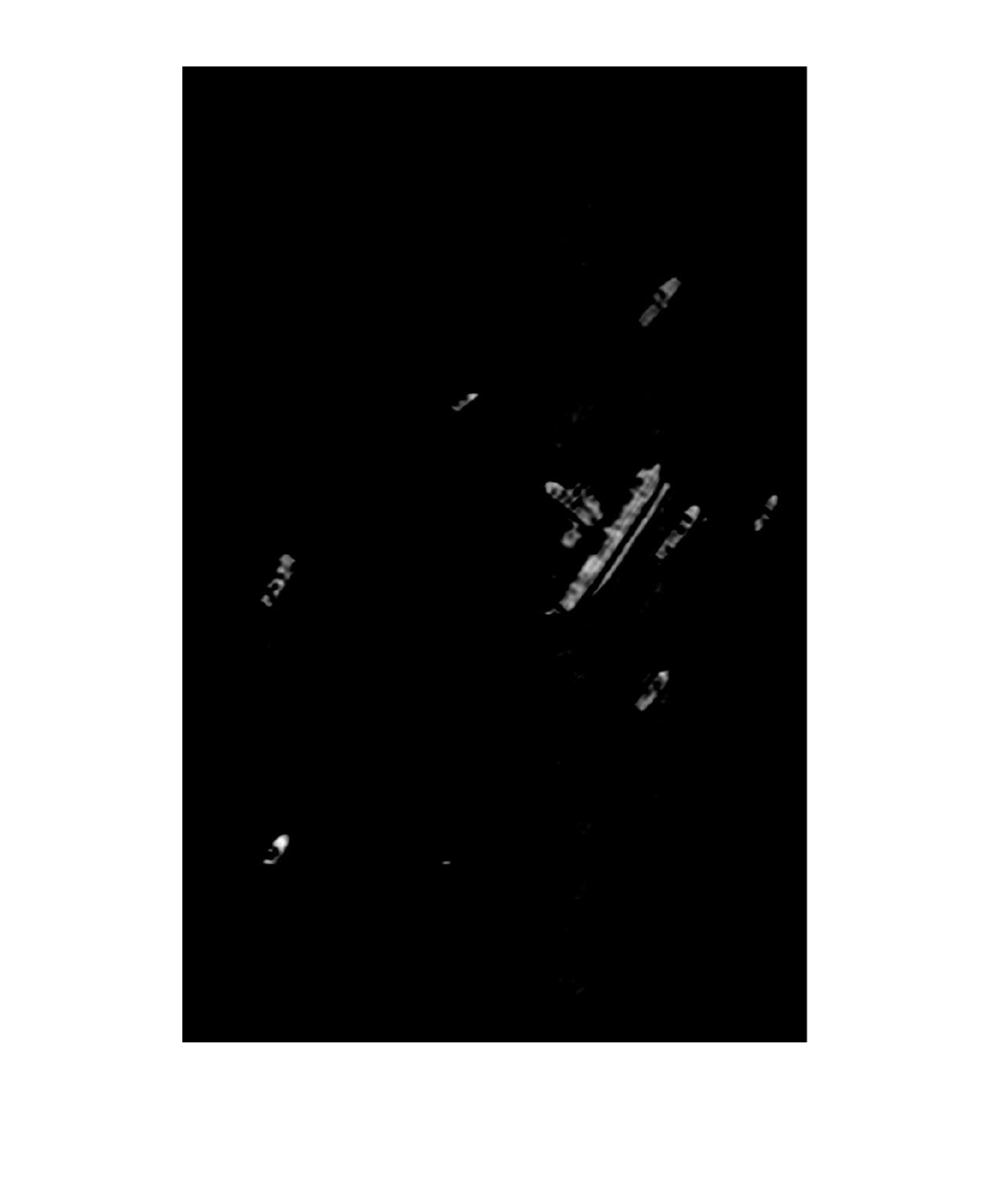} \\
(b)
\end{tabular}
\caption{Sub-Nyquist range sampling and recovery comparison. (a) Conventional RDA with full Nyquist samples. (b) SAR FISTA (Algorithm \ref{alg:SARFISTA}), $\bbpsi = I$, using $24\%$ of the coefficients required in conventional RDA.}
\label{fig:simulationSparse}
\end{figure}

In the first simulation we examined range subsampling, where the scene includes a sea with several vessels. Since there is nearly no back reflection from the water surface, large areas in the scene have almost no reflectivity, rendering the image spatially sparse. The number of transmitted pulses is $P=1200$ and the rest of the system parameters are described in Table~\ref{tab:SARparameters}. We processed the image using only $120$ Fourier coefficients from each received signal, instead of the $F_sT = 500$ which are required in order to satisfy RDA with $\os=2$. We compared conventional Range-Doppler processing with full samples, to Algorithm \ref{alg:SARFISTA}. The algorithm parameters are: $\beta=0.9, \lambda = 1000, \bar{\lambda}=1000$, $L_f = 1$ and $\bbpsi$ is taken as the identity transform. Figure~\ref{fig:simulationSparse}(a) depicts the scene processed with conventional RDA. Figure~\ref{fig:simulationSparse} (b) shows the result of our sub-Nyquist sampling and processing approach using Algorithm~\ref{alg:SARFISTA}. Our CS algorithm outperforms conventional RDA with only $24\%$ of the coefficients required in conventional RDA. The reconstructed image is sharper due to the attenuation of PSF sidelobes caused by the soft thresholding operation \eqref{eq:soft}.

\begin{figure*}
\centering
\begin{tabular}{cc}
\includegraphics[width=0.225\linewidth]{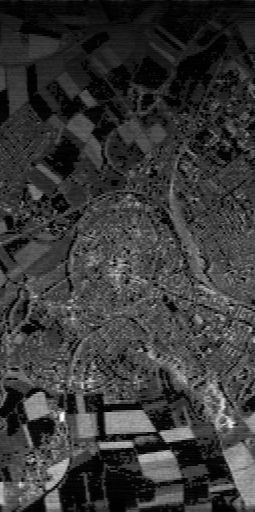}\;\;\;\;\;\;\;\;&  \;\;\;\;\;\;\;\;\;\;\;\;\;\;
\includegraphics[width=0.45\linewidth]{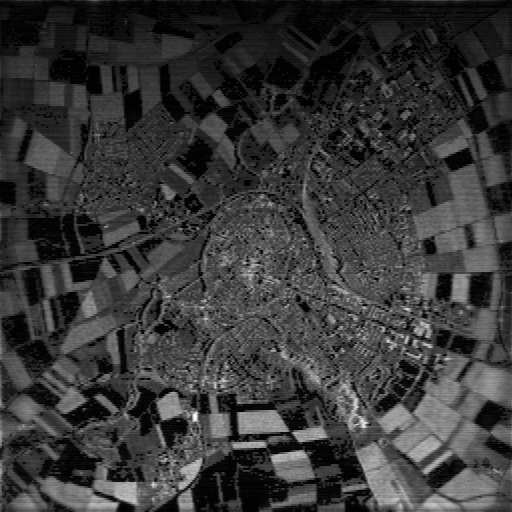}\\
(a)\;\;\;\;\;\;\;\;& \;\;\;\;\;\;\;\;\;\;\;\;\;\;(b)

\end{tabular}
\caption{Reduced time on scene via sub-Nyquist azimuth sampling (range is the vertical axis and azimuth is the horizontal one). (a) Conventional RDA with full Nyquist samples, 600 pulses. (b) Sub-Nyquist reconstruction of a 2 times wider scene using 600 pulses. The $50\%$ rate reduction enables to transmit the missing pulses to another area. The reconstruction is performed by SAR FISTA (Algorithm~\ref{alg:SARPRFFISTA}), where $\bbpsi$ is the Daubechies-4 wavelet transform.}
\label{fig:simulationSparseWavelets}
\end{figure*}

In the second simulation we examined only azimuth subsampling and demonstrated the reduced time on scene concept using the same system parameters as in the previous simulation, with full Nyquist sampling in range. The image includes two islands, and is not spatially sparse. Thus, we used the Daubechies-4 wavelet transform as the sparsifying basis. The PRF is dictated by the Nyquist theorem and should be higher than $25$ KHz for $l_a=6$ m. Thus, for a PRF of 30 KHz, the number of required pulses for a CPI of 20 msec is $P=600$. Figure~\ref{fig:simulationSparseWavelets}(a) shows the resulting image using conventional RDA with full Nyquist samples. In the second experiment we processed the data with Algorithm~\ref{alg:SARPRFFISTA} were only $300$ random pulses were chosen, instead of the required 600. Following the reduced time on scene concept we exploit the other 300 pulses in order to catch a wider part of the original scene, which means that using the same number of pulses we doubled the area of the captured image. The algorithm parameters are: $\beta=0.8, \lambda = 800, \bar{\lambda}=800$, $L_f = 1$. The result is shown in Fig.~\ref{fig:simulationSparseWavelets}(b). It can seen that using the same amount of pulses, our CS algorithm achieves results which are equivalent in terms of quality to the conventional RDA but the coverage is two times the original area. This result proves the concept of reduced time-on-scene.

\subsection{Real Data}
To further test the performance of our method and to confirm our model, we conducted simulations on the RADARSAT-1 raw data collected on June 16, 2002, in ascending orbit \#34522. The illuminated target is Richmond, Vancouver, Canada. The related key parameters of RADARSAT-1 system can be found in \cite{SARdspCumming}. In this simulation we prove the feasibility of our two-dimensional sub-Nyquist SAR system.

\begin{figure*}
\centering
\begin{tabular}{cc}
\multicolumn{2}{c}{
\includegraphics[width=0.8\columnwidth]{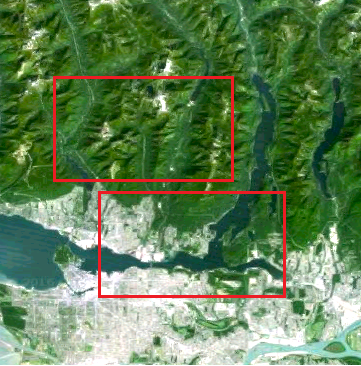}}\\
\multicolumn{2}{c}{(a)}\\
\includegraphics[width=\columnwidth]{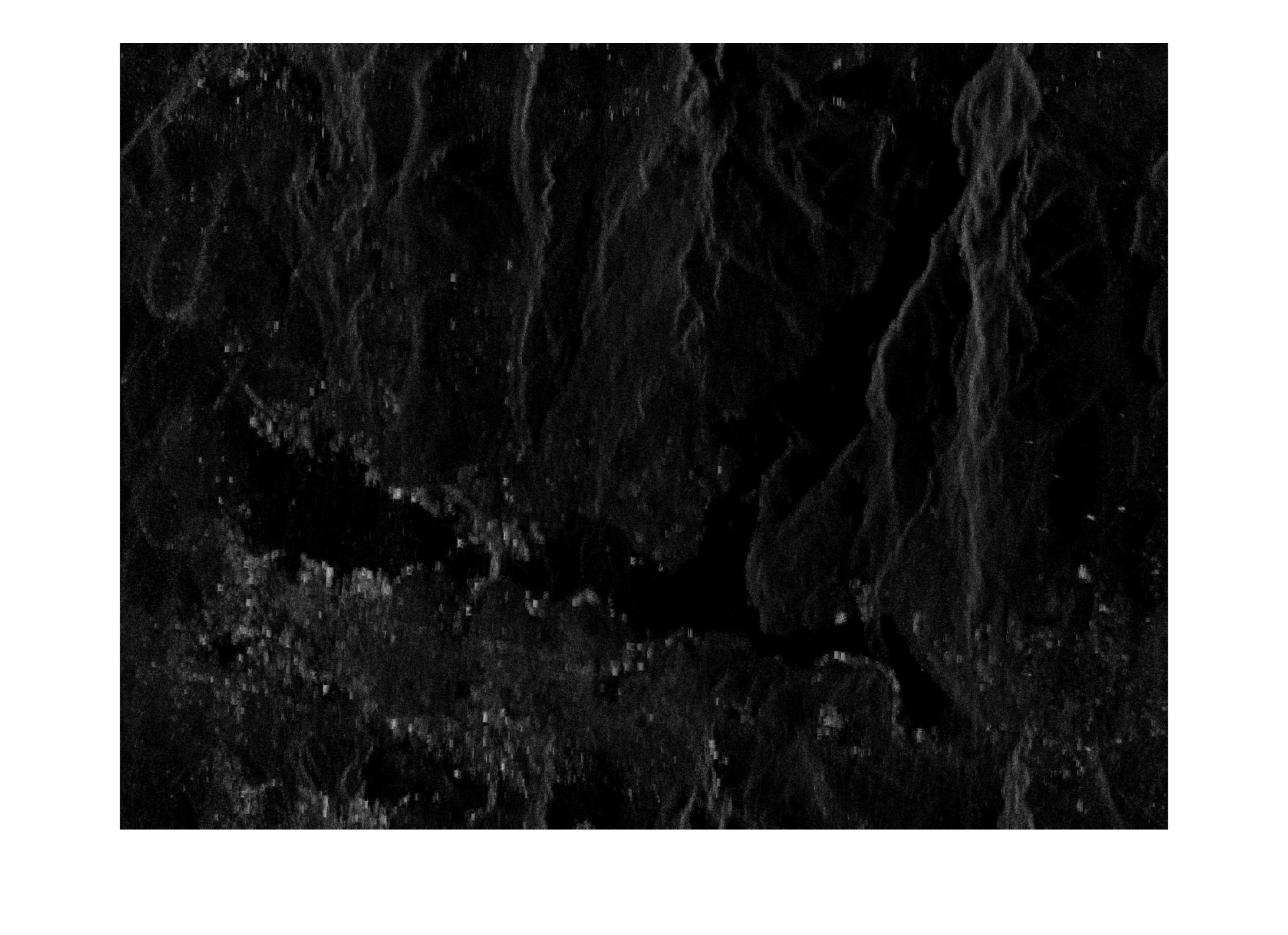} & \includegraphics[width=\columnwidth]{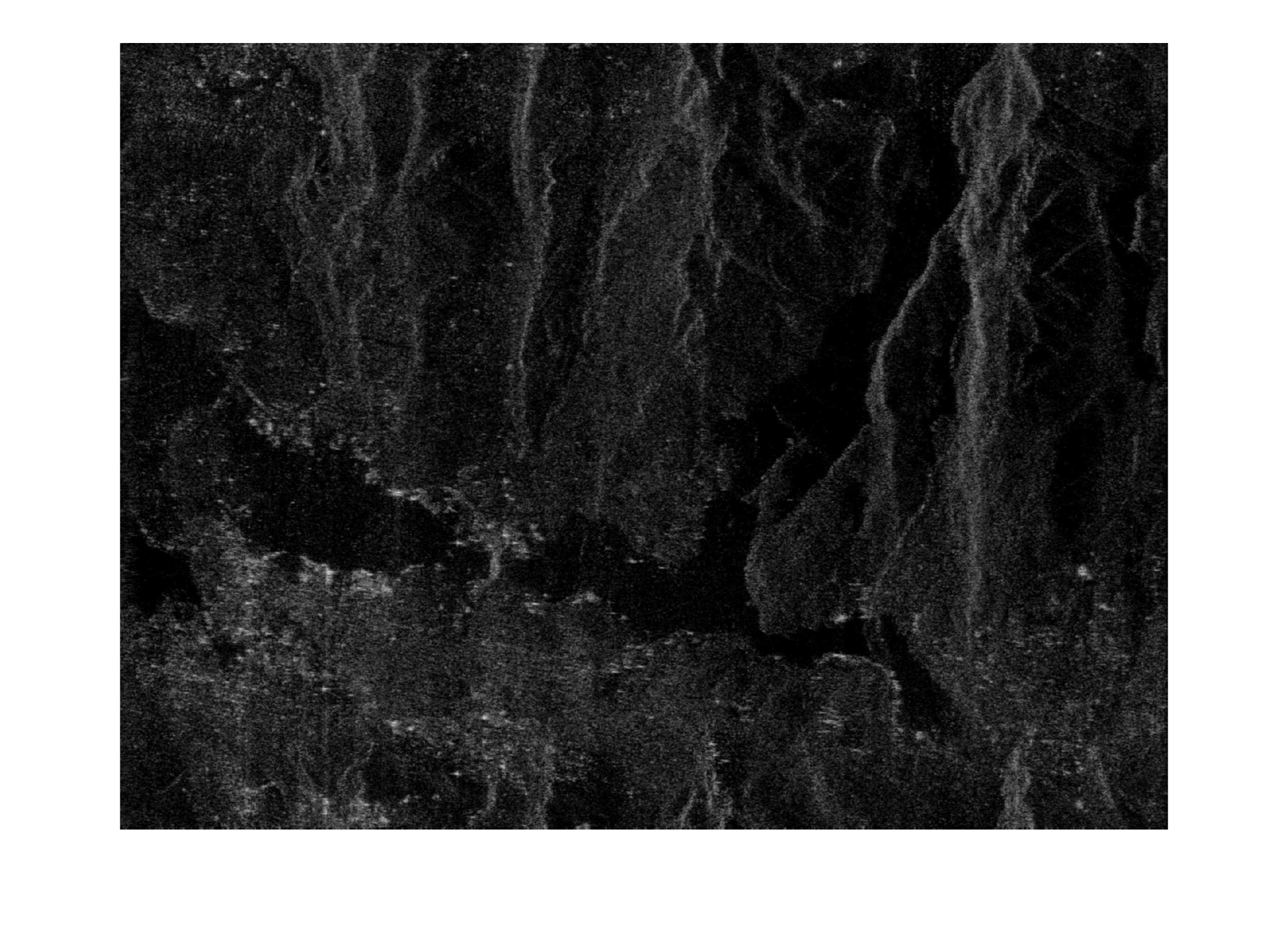}\\
(b) & (c) \\
\includegraphics[width=\columnwidth]{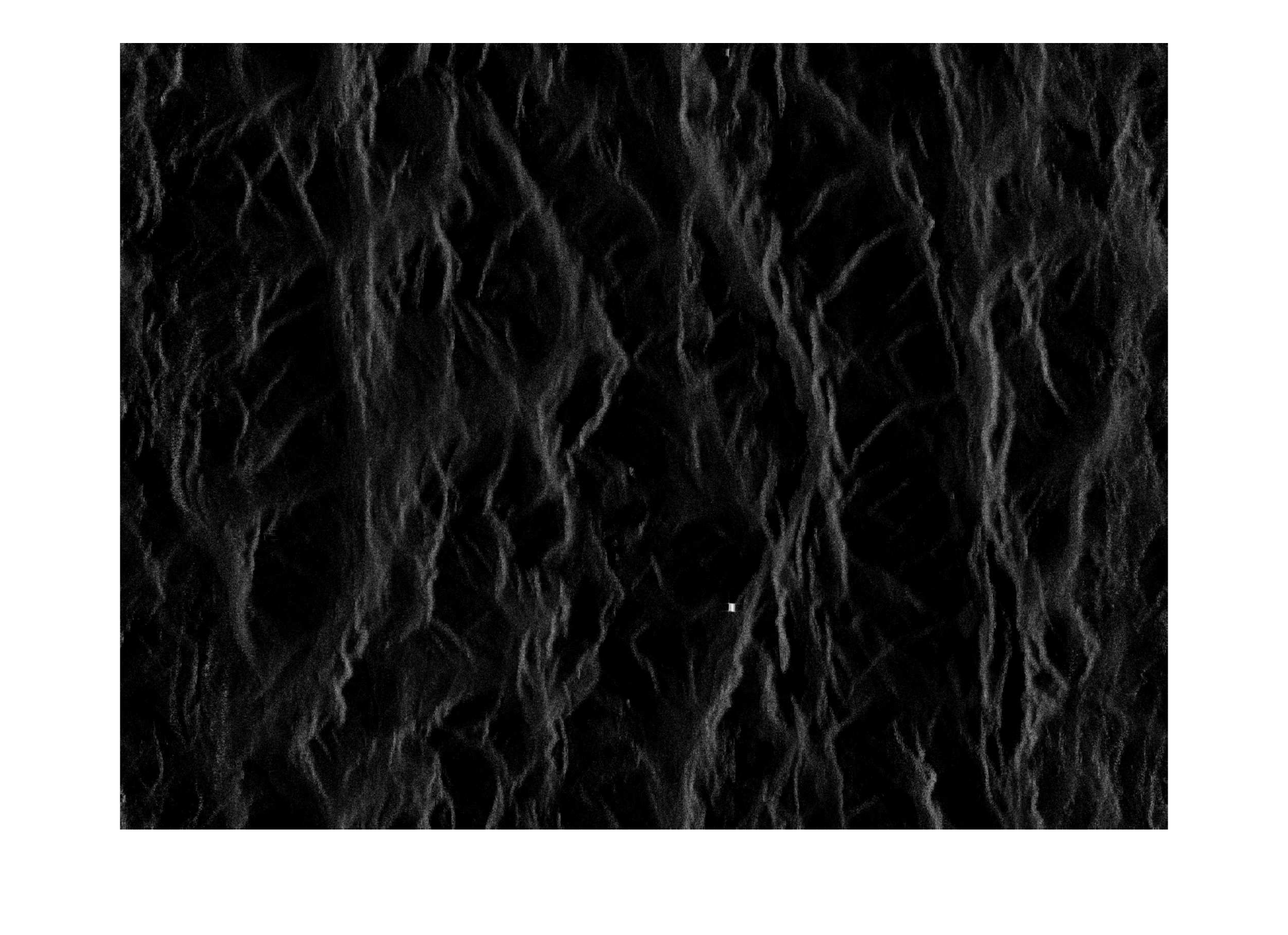} & \includegraphics[width=\columnwidth]{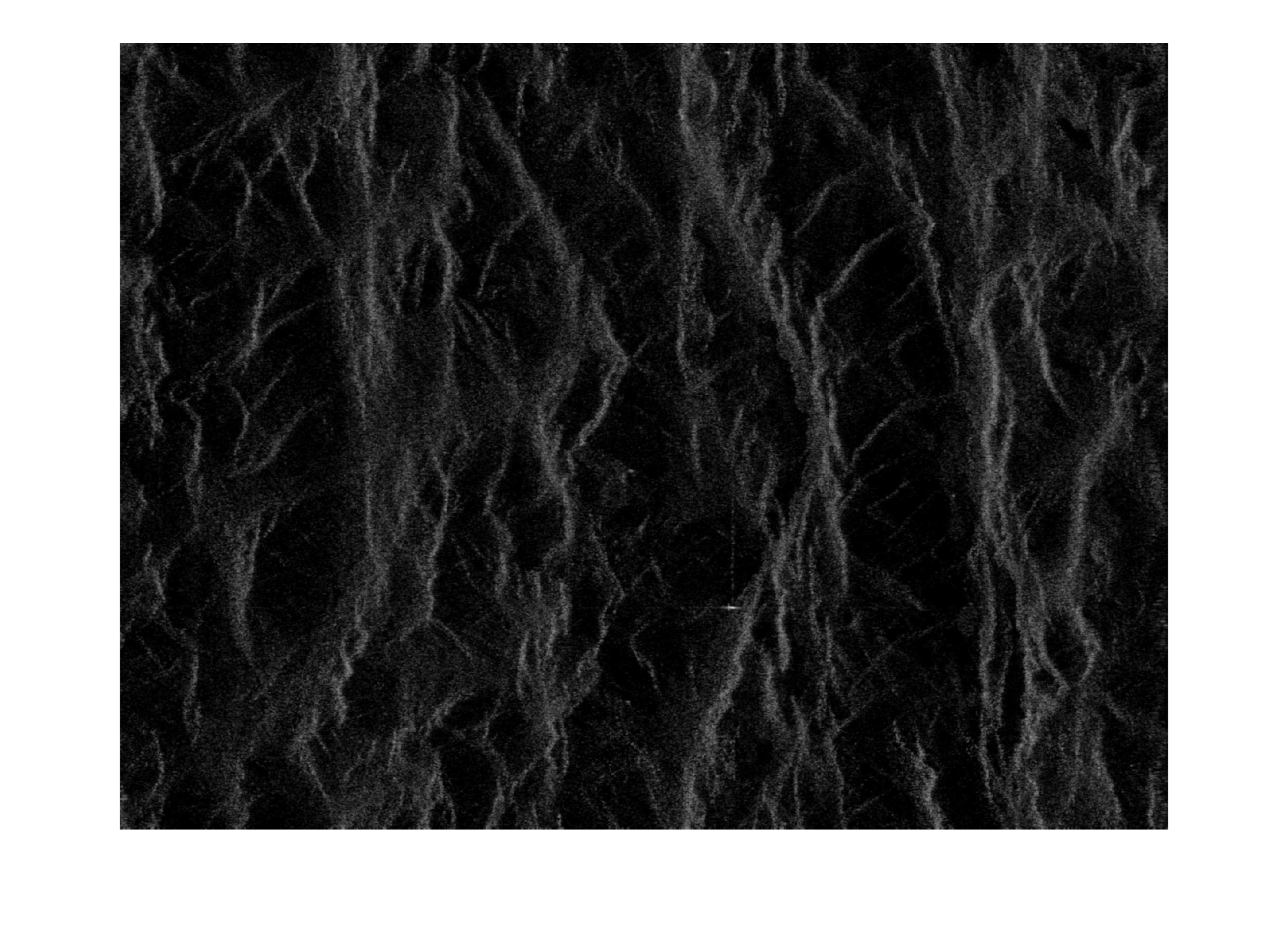} \\
(d) & (e) \\
\end{tabular}
\caption{Real data simulations using RADARSAT-1 data. (a) A reference electro-optic image with two marked areas (in red). (b) The first area (North), processed with full Nyquist samples. (c) The first area processed with Algorithm \ref{alg:SARPRFFISTA} using only 49\% of the original samples, 0.7 rate reduction in each axis. (d) The second area (South), processed with full Nyquist samples. (c) The second area processed with Algorithm \ref{alg:SARPRFFISTA} using only 49\% of the original samples, 0.7 rate reduction in each axis. }
\label{fig:realData}
\end{figure*}

We simulated two geographically consecutive scenes. The reference image which was taken from an Electro-Optic (EO) satellite is shown in Fig.~\ref{fig:realData}(a), where the two illuminated scenes are marked in red boxes.
In order to simulate the sub-Nyquist system we undersampled the original raw data of each of the scenes. The $3072\times 4096$ matrix was undersampled in both axes. In the range axis we reduced randomly 30\% of the coefficients and in the azimuth direction we selected randomly 70\% of the columns, which is equivalent to the omission of 30\% of the pulses. This leads to a $2150\times 2867$ reduced matrix. Then we use Algorithm \ref{alg:SARPRFFISTA} in order to reconstruct the images, using the Daubechies-4 wavelet transform as the sparsifying basis. The algorithm parameters were chosen as $\beta=0.9, \lambda = 0.01, \bar{\lambda}=0.001$, $L_f = 1$ and $\bbpsi$ is taken as the Daubechies-4 wavelet transform. The results in Fig.~\ref{fig:realData} compare the original processing with full Nyquist samples and the sub-Nyquist recovery method using only $0.7^2 = 0.49$ of the original samples in each image. It can be readily seen that despite the missing samples, the detailed images are well reconstructed. In practice, when we control the transmitted signal, we can increase the signal's power at the subsampled bands and increase the effective SNR for better results, as seen in the next sub-section.

\begin{table}
\caption{Simulated SAR system parameters}
\label{tab:SARparameters}
\centering
\bgroup
\def\arraystretch{1.5}
\begin{tabular}{l c}
\hline
SAR parameter & Value\\
\hline\hline Carrier frequency -- $f_c$ & 37.5 MHz\\
Transmitted pule duration -- $\tau$ & 1.67 $\mu$s\\
Chirp rate -- $K_r$ & 2.25 GHz/msec\\
 Range sampling rate -- $F_s$ & 11.25 MHz \\
 Sensor velocity -- $v$& 75 km/s\\
 Doppler bandwidth & 25 KHz\\
 PRF -- $1/T$ & 30 KHz\\
Squint angle & $0^{\circ}$\\
\hline
\end{tabular}
\egroup
\end{table}

\subsection{Cognitive Radar on Hardware}
\label{sec:experimental}
In order to demonstrate our cognitive SAR abilities, we now present a real experiment of our SAR hardware prototype. We integrate our method into a stand-alone system and show that such processing is feasible in practice using real hardware. Our setup includes a custom made sub-Nyquist receiver board which implements sub-Nyquist Xampling and digital recovery using Algorithm~\ref{alg:SARPRFFISTA}, while the analog input signal \eqref{eq:received} was synthesized using National Instruments (NI) hardware.

The experimental process consists of the following steps. We begin by using the AWR software, which enables us to simulate point reflectors with different amplitudes and spatial distribution.

With the AWR software we simulate the complete radar scenario, including the pulse transmission and accurate power loss due to wave propagation in a realistic medium. The AWR also contains a model of a realistic RF receiver, which simulates the demodulation of the RF signal to IF frequencies, and saves the output to a file. However, since AWR is operated only with stationary radars, in order to simulate SAR signals we created an equivalent kinematic state. We simulated the targets with velocity $v$, but in the opposite direction than the one that should be to the radar. Our simulation is similar in some manners to ISAR. In ISAR, the radar is stationary and the targets are moving. The angular motion of the target with respect to the radar can be used to form an image of the moving targets. Differential Doppler shifts of adjacent scatters on a target are observed and the target’s reflectivity function is obtained through the Doppler frequency spectrum \cite{wehner1987high}.

\begin{figure}
\centering
\begin{tabular}{c}
\includegraphics[width=0.75\linewidth]{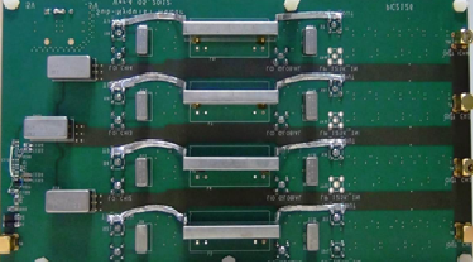} \\ (a) \\
\includegraphics[width=0.75\linewidth]{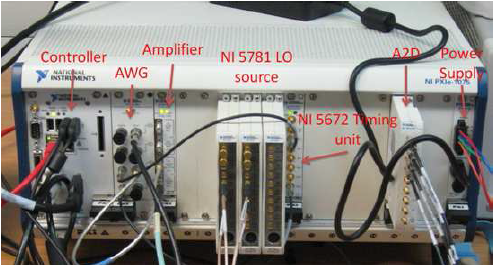}  \\
(b)\\
\end{tabular}
\caption{Sub-Nyquist system. (a) Analog front end 4-channel receiver board. (b) National Instruments (NI) system.}
\label{fig:hardware}
\end{figure}

Next, the generated raw data is loaded to the AWG module, which produces an analog signal. This signal is amplified using the NI 5690 low noise amplifier and then routed to our radar receiver board, which has 4 parallel input channels. Each channel samples a different frequency band, in the following manner: each channel is fed by a local oscillator (LO), which modulates the desired frequency band of the channel to the central frequency of a narrow 80 KHz bandwidth band pass filter (BPF). A fifth LO, common to all 4 channels, modulates the BPF output to a low frequency band. It is then sampled with a standard low rate ADC. The LOs are created using three NI 5781 baseband transceivers, acting as trigger based signal generators with a constant and known phase, controlled by NI Flex Rio FPGAs. The AWG also triggers the ADC to sample 250 samples in each sampling cycle, per channel. These samples are fed into the chassis controller and a MATLAB function is launched that runs Algorithm~\ref{alg:SARPRFFISTA}. Pictures of the system are shown in Fig.\ref{fig:hardware}.

\begin{figure}
\centering
\begin{tabular}{cc}
\includegraphics[width=0.5\columnwidth]{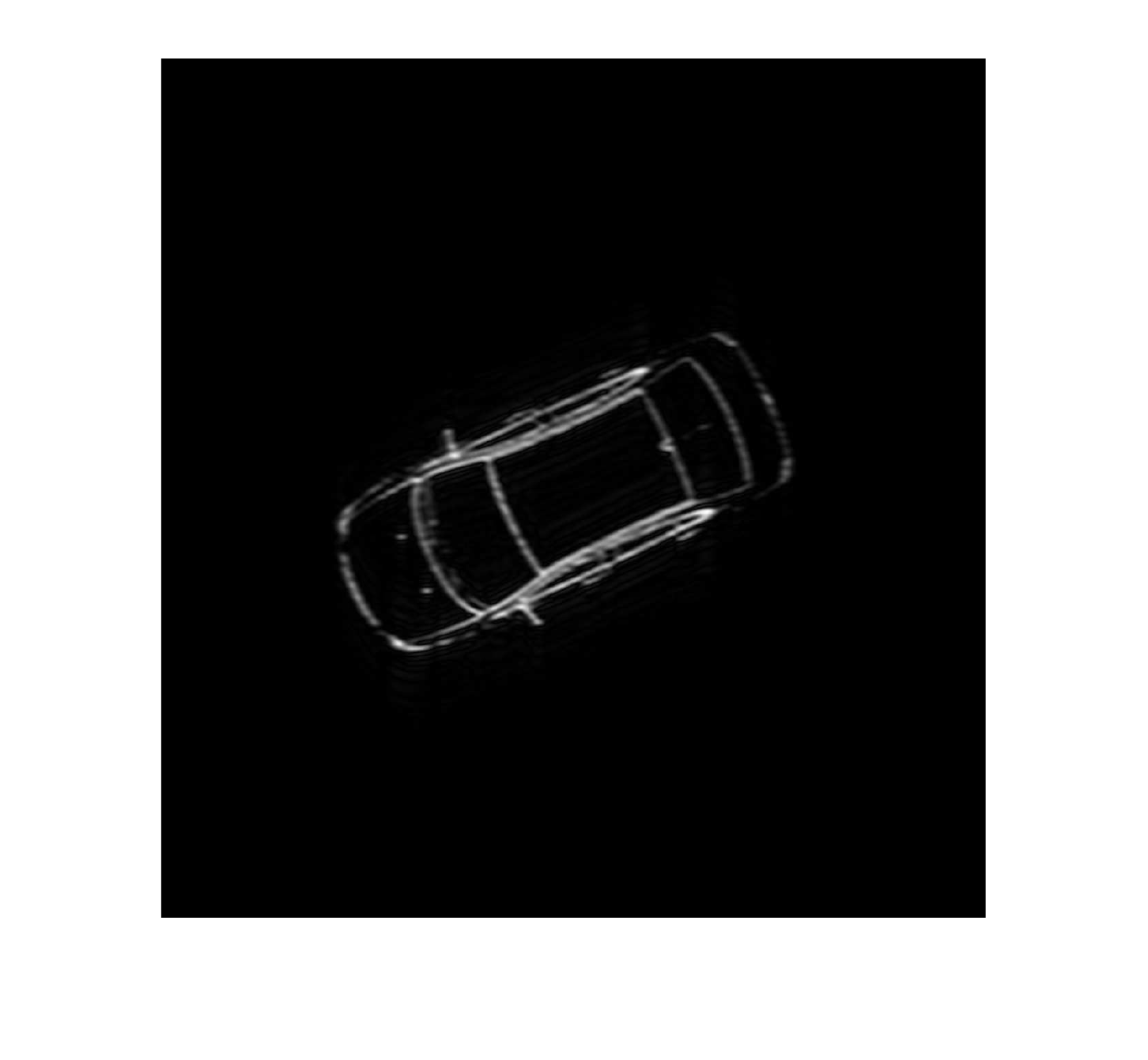} & \includegraphics[width=0.5\columnwidth]{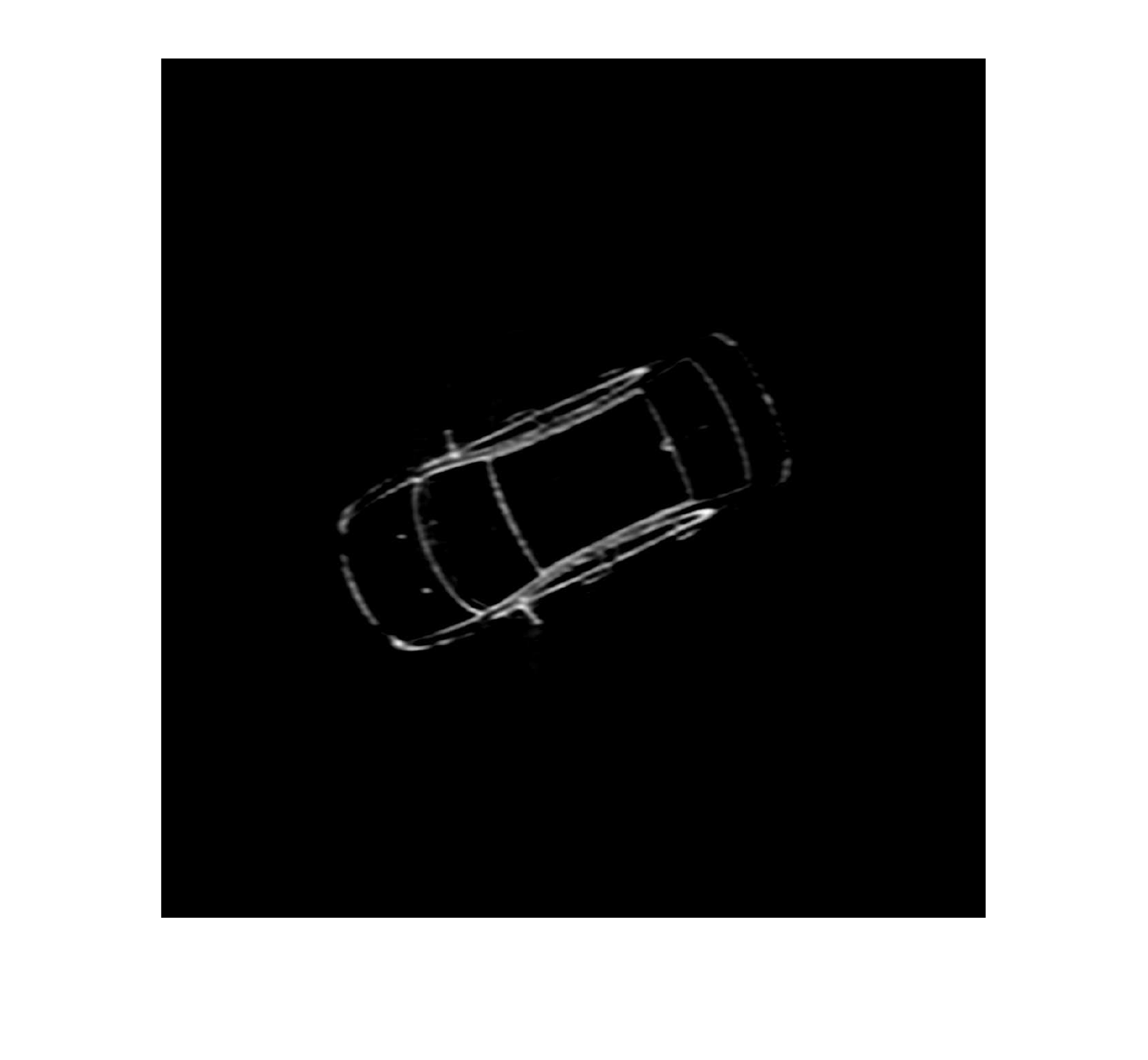} \\
(a) & (b) \\
\includegraphics[width=0.5\columnwidth]{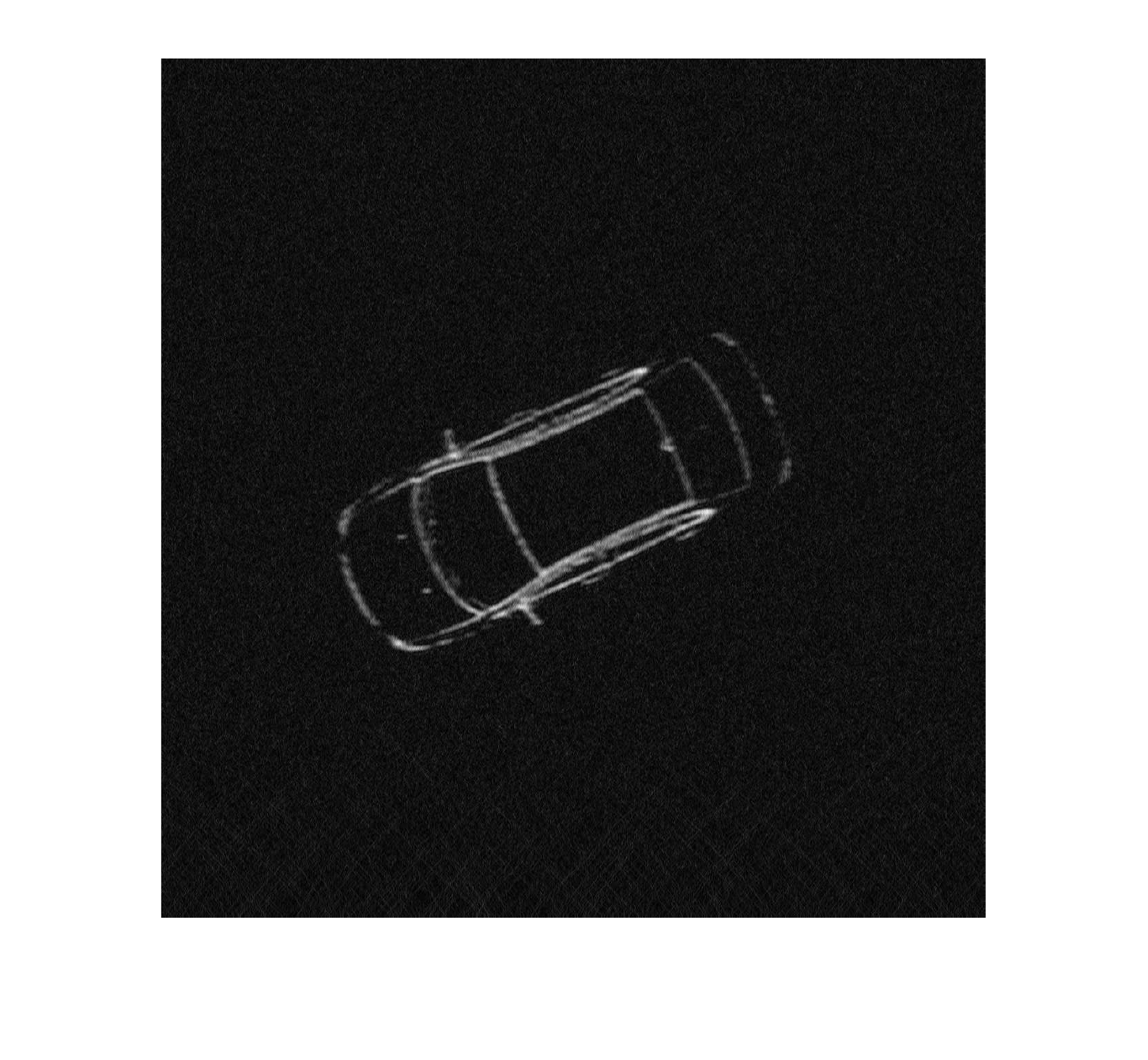} & \includegraphics[width=0.5\columnwidth]{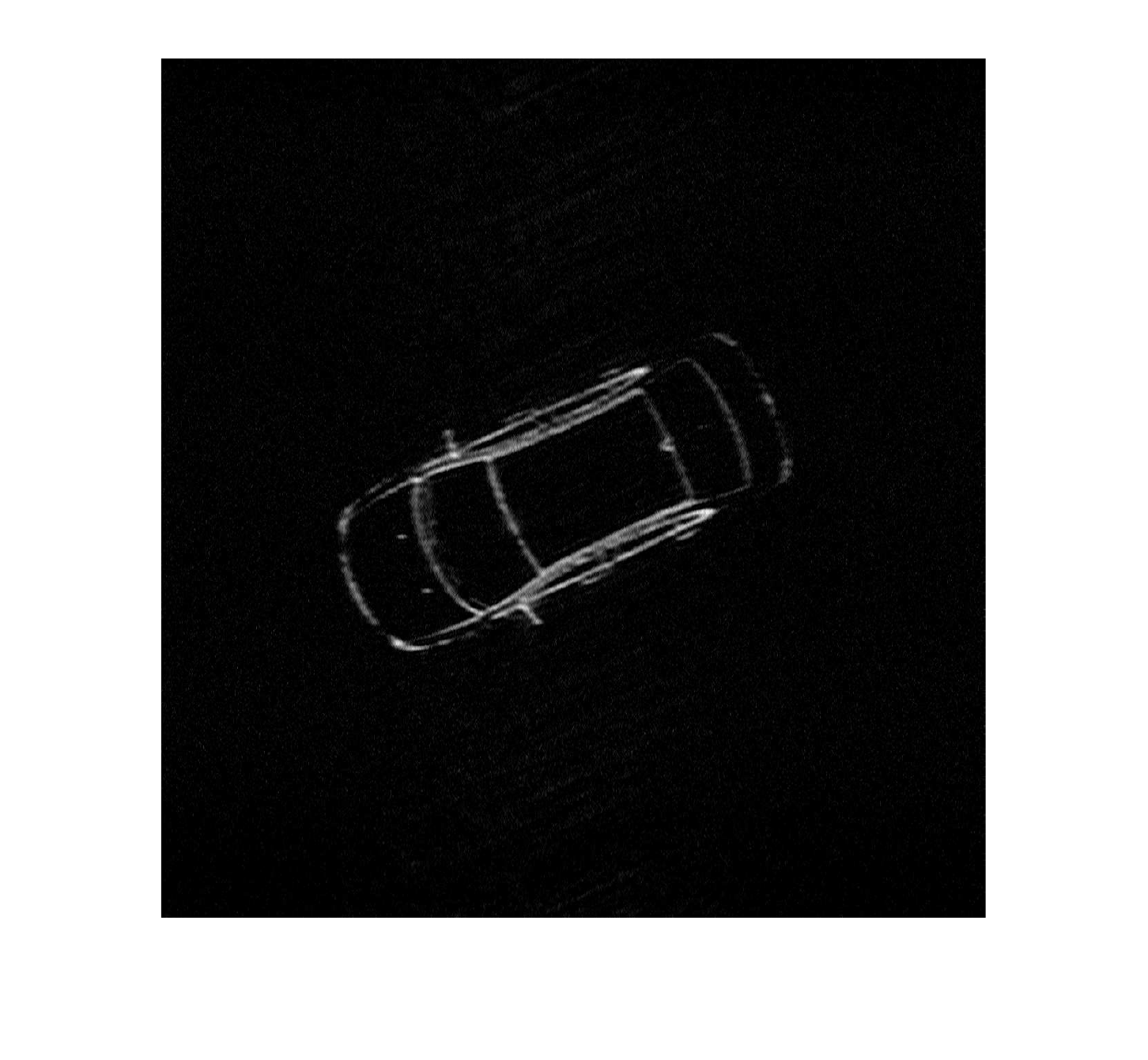} \\
(c) & (d) \\
 \includegraphics[width=0.5\columnwidth]{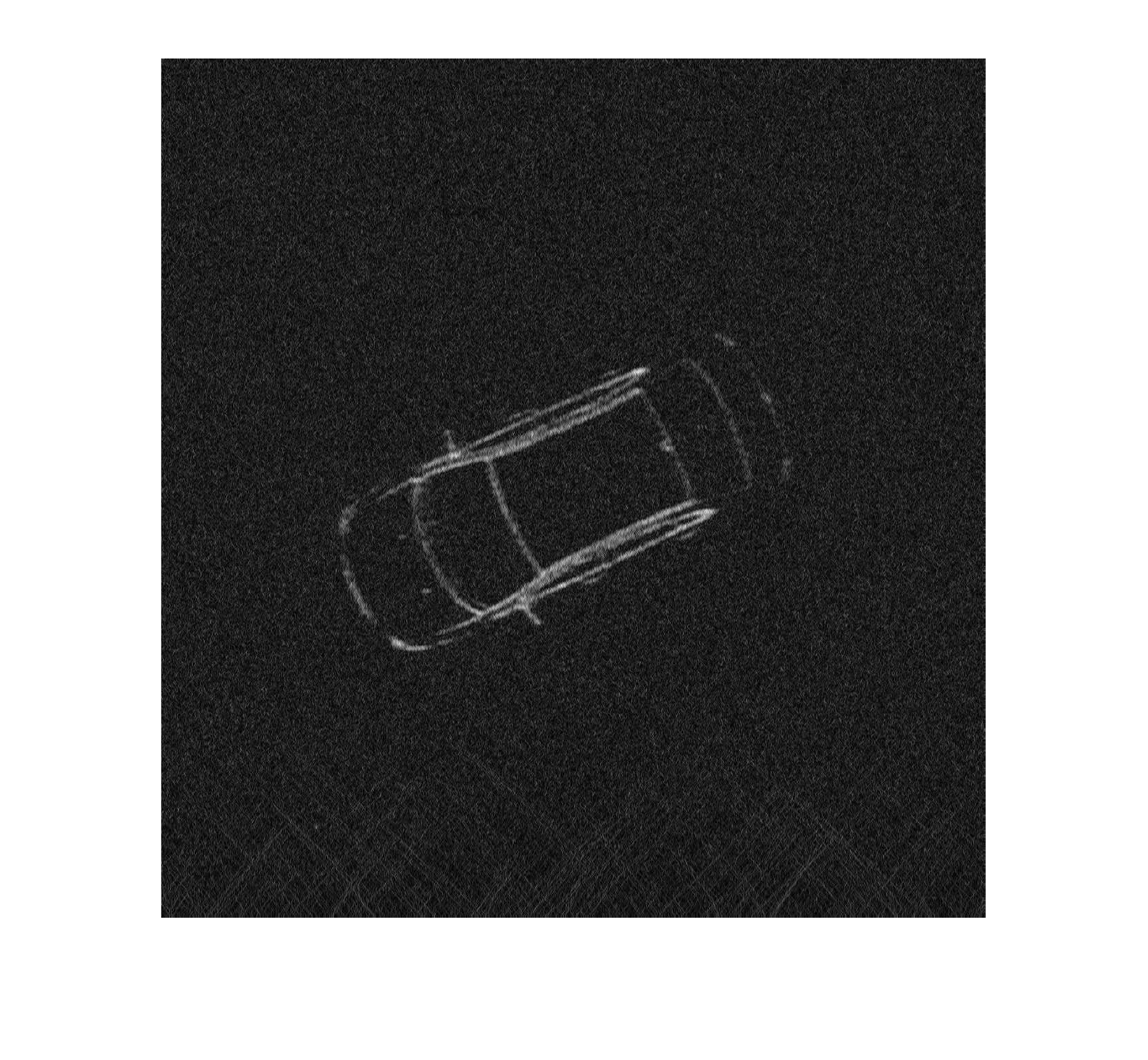} & \includegraphics[width=0.5\columnwidth]{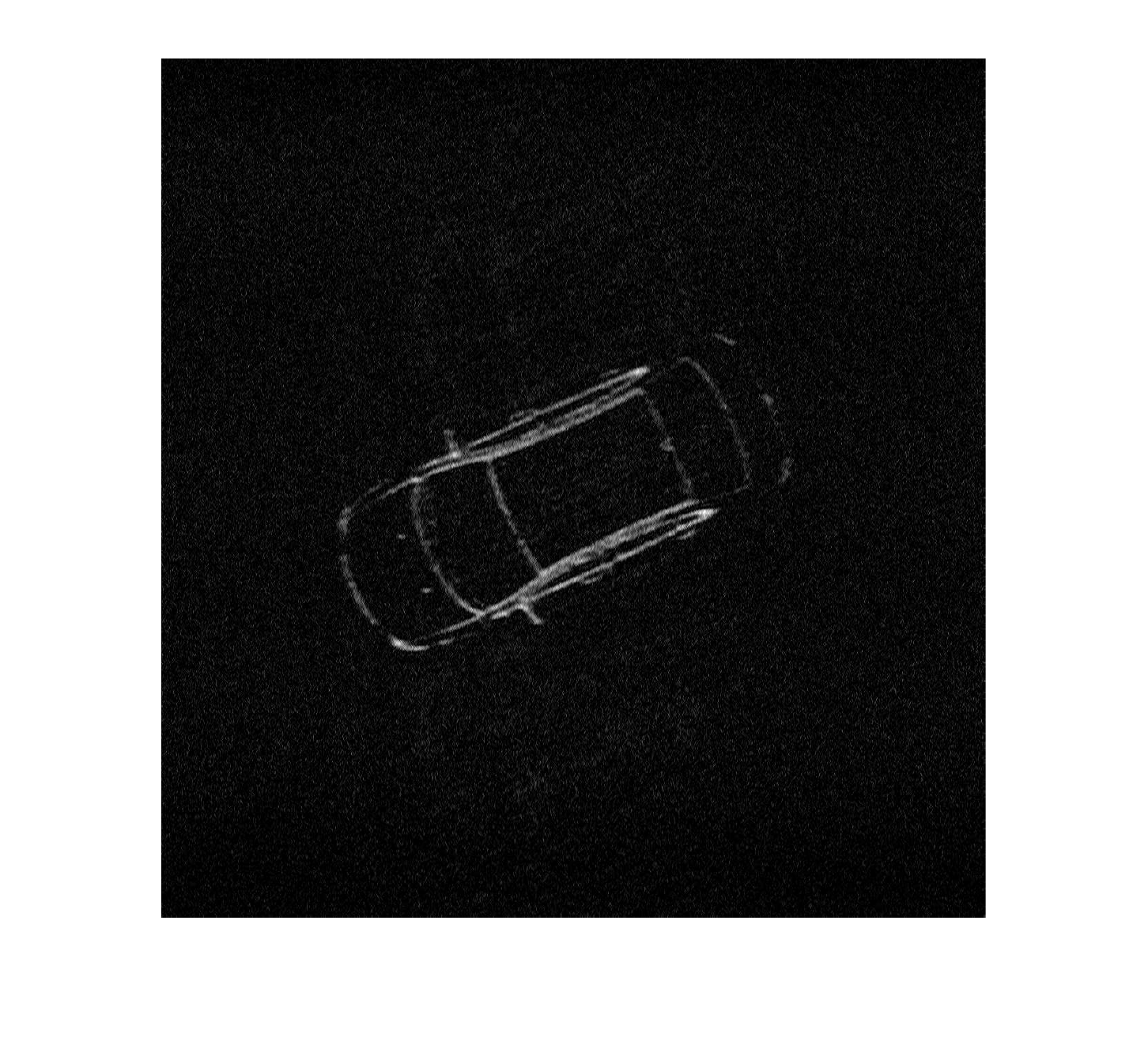} \\
(e) & (f)\\
\end{tabular}
\caption{SAR experiment using real hardware. Comparing between full-Nyquist noise free processing using conventional RDA and SAR FISTA using only 20\% of the samples. (a) Noise free, full-Nyquist samples, conventional RDA processing. FSIM = 1. (b) Noise free, SAR FISTA. FSIM = 0.9992. (c) SNR = -10 dB. Full-Nyquist samples, conventional RDA processing. FSIM = 0.881. (d) SAR FISTA, using 20\% of the original samples. FSIM = 0.994. (e) SNR = -20 dB. Full-Nyquist samples, conventional RDA processing. FSIM = 0.852. (f) SAR FISTA, using 20\% of the original samples. FSIM = 0.982.
}
\label{fig:carEdges}
\end{figure}

To demonstrate the advantage of our cognitive system in terms of SNR using hardware, we simulated targets which construct an ISAR frame of a moving car, which means that only the car edges can be detected, thus, the image is spatially sparse and $\bbpsi$ is selected to be the identity. We examined 3 scenarios with different SNR: noise free, -10 dB and -20 dB. In each scenario we ran two simulations: a fully sampled signal, according to that required by RDA, and a partially sampled signal with only $20\%$ of the coefficients, using our hardware. The algorithm parameters are: $\beta=0.85, \lambda = 0.001, \bar{\lambda}=0.0001$, $L_f = 25$. For every resulting image, we computed the FSIM index compared to the original full Nyquist reconstructed image without noise. Figure~\ref{fig:carEdges} compares the results. Due to the power concentration in the sampled bands which helps increase the effective SNR, the FSIM is larger in our method.

Next, we compared the quality of two images using FSIM: the output of traditional processing and the image resulting from our hardware implementing a cognitive system. The image includes a few single point reflectors which were randomly located. At every measurement each of the resulting images was compared to a ground truth noiseless image processed using conventional RDA. The received signals were corrupted with additive white Gaussian noise
(AWGN) $n(t)$ with power spectral density $N_0/2$, bandlimited to $B_h$. The SNR for a single reflector located at $\br_0$ is defined as
\begin{equation}
\label{eq:SNR}
\textrm{SNR} = \frac{\frac{1}{T}\int_0^T|\sigma(\br_0)h(t)|^2dt}{N_0B_h}.
\end{equation}
Figure~\ref{fig:FSIM} plots the FSIM as a function of SNR. The index values are in the range of 0 to 1, where 1 indicates perfect similarity. Evidently, our cognitive system, with a lower number of samples, outperforms traditional wideband radar transmission and processing.

\begin{figure}
\centering
\includegraphics[width= \linewidth]{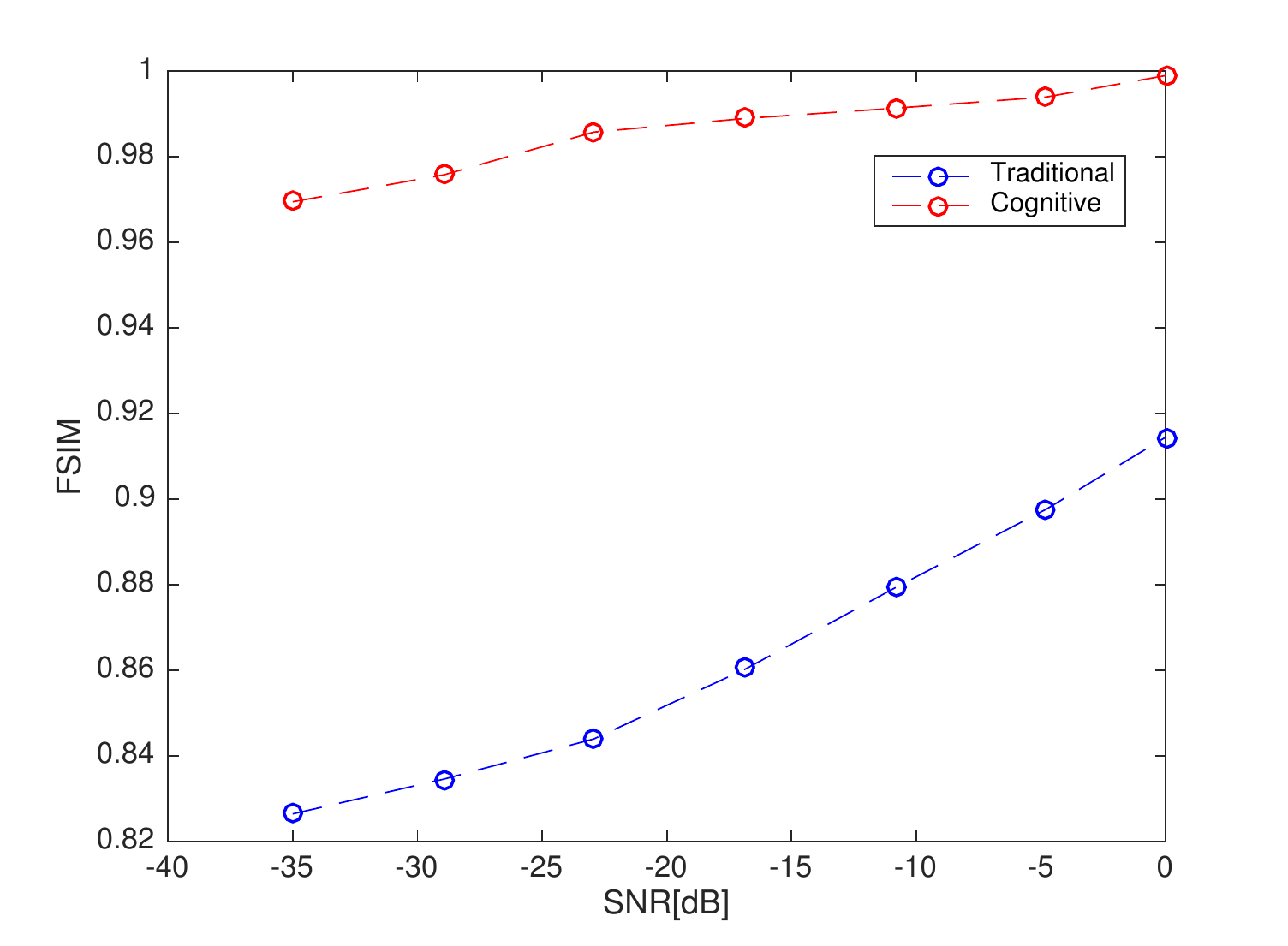}
\caption{Comparing between traditional and cognitive SAR images using the FSIM index.}
\label{fig:FSIM}
\end{figure}

Our experimental prototype proves that the sub-Nyquist methodology described in this paper is feasible in practice. The proposed recovery method addresses the problem of low rate analog sampling, in a way which is feasible with standard RF hardware. In addition, in terms of SNR, our algorithm outperforms conventional RDA while using only a portion of the original samples due to the fact the we concentrated the energy only in the sampled bands.

\section{Conclusion}
\label{sec:Conclusion}
We presented a new SAR signal processing algorithm which is equivalent to RDA and showed that the resulting images are equivalent to those of conventional processing. The new algorithm exploits the advantages of RDA without the heavy interpolation stage. This allows to perform processing at the Nyquist rate, defined with respect to the effective bandwidth of the signal, which is impossible when interpolation is performed in time.

Next, we introduced two-dimensional sub-Nyquist sampling and recovery methods, which employ the techniques of Xampling. We showed that an image can be reconstructed while sampling only a portion of its bandwidth and after dropping a large percentage of the transmitted pulses. The gaps in time and frequency may be exploited in order to achieve wider coverage during the same CPI, to increase SNR and to adapt the transmitted signal to the environment, paving the way to cognitive SAR. Using simulated and real data sets, and a Xampling prototype in hardware, we demonstrated that our system outperforms conventional SAR and can cope with practical limitations of computational load and limited bandwidth.

\section*{Acknowledgment}
The authors would like to thank the anonymous reviewers for their constructive remarks, which helped improve the presentation.

\bibliographystyle{IEEEtran}
\bibliography{../my_references}

\end{document}